\documentclass[twocolumn,showpacs,aps,prd,superscriptaddress]{revtex4}

\usepackage{graphicx}
\usepackage{dcolumn}
\usepackage{amsmath}
\usepackage{epsfig}
\usepackage{enumerate}

\RequirePackage{xspace}

\input babarsym

\newcommand{\BABARPubYear}    {08}
\newcommand{\BABARPubNumber}  {027}

\newcommand{\SLACPubNumber} {13371}

\def\DSlope     {1.20}
\def\DSlopeStE  {0.04}
\def\DSlopeSyE  {0.07}
\def\DsSlope    {1.22}
\def\DsSlopeStE {0.02}
\def\DsSlopeSyE {0.07}
\def\DBF        {2.34}
\def\DBFStE     {0.03}
\def\DBFSyE     {0.13}
\def\DsBF       {5.40}
\def\DsBFStE    {0.02}
\def\DsBFSyE    {0.21}
\def\GVcb       {43.1}
\def\GVcbStE    {0.8}
\def\GVcbSyE    {2.3}
\def\FVcb       {35.9}
\def\FVcbStE    {0.2}
\def\FVcbSyE    {1.2}
\def\corGSlope  {+0.64}
\def\corFSlope  {+0.56}
\def\corGF      {-0.07}
\def\VcbG       {39.9}
\def\VcbGStE    {0.8}
\def\VcbGSyE    {2.2}
\def\VcbGThE    {0.9}
\def\VcbF       {38.6}
\def\VcbFStE    {0.2}
\def\VcbFSyE    {1.3}
\def\VcbFThE    {1.0}
\def\GFRatio    {1.20}
\def\GFRatioE   {0.09}

\def\cosBDl {\ensuremath{\cos\theta_{B-D\ell}}\xspace}

\def\figurebox#1#2#3{%
    \def\arg{#3}%
    \ifx\arg\empty
    {\hfill\vbox{\hsize#2\hrule\hbox to #2{\vrule\hfill\vbox to #1{\hsize#2\vfill}\vrule}\hrule}\hfill}%
    \else
    {\hfill\epsfbox{#3}\hfill}%
    \fi}

\begin{document}

\preprint{\babar-PUB-\BABARPubYear/\BABARPubNumber} 
\preprint{SLAC-PUB-\SLACPubNumber} 

\begin{flushleft}
\babar-PUB-\BABARPubYear/\BABARPubNumber\\
SLAC-PUB-\SLACPubNumber\\
\end{flushleft}

\title{
{\large \bf
Measurements of the Semileptonic Decays
$\Bb\to D\ell\nub$ and $\Bb\to D^*\ell\nub$ 
Using a Global Fit to $D X\ell\nub$ Final States} 
}

%
\author{B.~Aubert}
\author{M.~Bona}
\author{Y.~Karyotakis}
\author{J.~P.~Lees}
\author{V.~Poireau}
\author{E.~Prencipe}
\author{X.~Prudent}
\author{V.~Tisserand}
\affiliation{Laboratoire de Physique des Particules, IN2P3/CNRS et Universit\'e de Savoie, F-74941 Annecy-Le-Vieux, France }
\author{J.~Garra~Tico}
\author{E.~Grauges}
\affiliation{Universitat de Barcelona, Facultat de Fisica, Departament ECM, E-08028 Barcelona, Spain }
% ...
\collaboration{The \babar\ Collaboration}
\noaffiliation

\date{\today}

\begin{abstract}
Semileptonic $\Bb$ decays to $D X\ell\nub$ ($\ell=e$ or $\mu$) are
selected by reconstructing $\Dz\ell$ and $\Dp\ell$ combinations from a
sample of 230 million $\FourS\to\BB$ decays recorded with the \babar\
detector at the \pep2 $\epem$ collider at SLAC.  A global fit to these
samples in a 3-dimensional space of kinematic variables is used to
determine the branching fractions $\mathcal{B}(\Bm\to D^0\ell\nub) =
(\DBF \pm \DBFStE\pm \DBFSyE) \%$ and $\mathcal{B}(\Bm\to D^{*0}\ell\nub) = 
(\DsBF\pm \DsBFStE\pm \DsBFSyE) \%$
where the errors are statistical and systematic, respectively.  
The fit also determines form factor parameters in a HQET-based
parameterization, resulting in $\rho^2_D=\DSlope\pm \DSlopeStE\pm \DSlopeSyE$ for $\Bb\to
D\ell\nub$ and $\rho^2_{\Dstar}=\DsSlope\pm \DsSlopeStE\pm \DsSlopeSyE$
for $\Bb\to\Dstar\ell\nub$.  These values are
used to obtain the product of the CKM matrix element \Vcb\ times the
form factor at the zero recoil point for both $\Bb\to D\ell\nub$
decays, $\mathcal{G}(1)\Vcb=(\GVcb\pm \GVcbStE\pm \GVcbSyE)\times10^{-3}$, and for $\Bb\to \Dstar\ell\nub$
decays, $\mathcal{F}(1)\Vcb=(\FVcb\pm \FVcbStE\pm \FVcbSyE)\times10^{-3}$.
\end{abstract}


\maketitle

\setcounter{footnote}{0}

\section{Introduction}
\label{sec:Introduction}
The study of semileptonic decays of heavy quarks provides the cleanest
avenue for the determination of several elements of the
Cabibbo-Kobayashi-Maskawa matrix~\cite{CKM}, 
which are fundamental parameters in
the standard model of particle physics.  The coupling strength of the
weak $\b\to\c$ transition is proportional to \Vcb, which
has been measured in both
inclusive semileptonic $\B$ decays~\cite{ref:VcbInclMeas} and in the
exclusive transitions 
$\Bb\to D\ell\nub$~\cite{ref:ALEPHDstarlnu,ref:CLEODlnu,ref:BelleDlnu,ref:BaBarbclnutagged} and
$\Bb\to\Dstar\ell\nub$~\cite{ref:ALEPHDstarlnu,ref:CLEODstarlnu,
ref:BelleDstarlnu,ref:BaBarDstarlnu,ref:BaBarbclnutagged,ref:BaBarDstarzlnu}
($\ell=e$ or $\mu$ and charge conjugate modes are implied).  
The inclusive and exclusive determinations
of \Vcb\ rely on different theoretical calculations. 
The former
employs a parton-level calculation of the decay rate organized in a
double expansion in $\alpha_S$ and in inverse powers of $m_b$, the
$\b$-quark mass.
The latter relies on a parameterization of the
decay form factors using Heavy Quark Symmetry and a non-perturbative
calculation of the form factor normalization at the zero recoil
(maximum squared momentum transfer) point.  The theoretical
uncertainties in these two approaches are independent.  The inclusive and
exclusive experimental
measurements use different techniques and have negligible
statistical overlap, and thus have largely uncorrelated 
uncertainties.  This
independence makes the comparison of \Vcb\ from inclusive and
exclusive decays a powerful test of our understanding of semileptonic
decays.  The latest determinations~\cite{ref:PDG2008} differ by
more than two standard deviations ($\sigma$), and the inclusive
determination is currently more than twice as precise as the
exclusive determination.  
Improvements in the measurements of exclusive decays will strengthen
this test.  This is particularly true
for the $\Bb\to D\ell\nub$ decay, where the experimental uncertainties
dominate the determination of $\Vcb$.
For the decay $\Bzb\to\Dstarp\ell\nub$, the experimental situation
needs clarification, as existing measurements are in poor agreement with each
other~\cite{ref:PDG2008}.  
Finally, precise measurements of semileptonic $\B$ decays to charm are
needed to further improve determinations of \Vub, where 
$\Bb\to D^{(*)}\ell\nub$ decays are 
the principal background.

Semileptonic $\b\to\c$ transitions result in the production of a charm
system that cascades down to the ground state $\Dz$ or $\Dp$ mesons.
Most previous analyses have focused
on reconstructing separately the exclusive decays
$\Bb\to\Dstar\ell\nub$~\cite{ref:ALEPHDstarlnu,ref:CLEODstarlnu,
ref:BelleDstarlnu,ref:BaBarDstarlnu,ref:BaBarDstarzlnu} 
and $\Bb\to D\ell\nub$~\cite{ref:ALEPHDstarlnu,ref:CLEODlnu,ref:BelleDlnu}.  
The $\Bb\to\Dstar\ell\nub$ analyses involve reconstruction of the soft
transition pion from the decay $\Dstar\to D\pi$, which is at the
limit of detector acceptance; determination of
the reconstruction efficiency for these pions
introduces significant systematic uncertainty.  Studies of the
exclusive decay $\Bb\to D\ell\nub$ suffer from large feed-down
background from $\Bb\to\Dstar\ell\nub$ decays where the transition
pion is undetected. 

In this analysis we reconstruct $\Dz\ell$ and $\Dp\ell$ pairs
and use a global fit to their kinematic properties to
determine the branching fractions and form factor parameters of the dominant
semileptonic decays $\Bb\to D\ell\nub$ and $\Bb\to\Dstar\ell\nub$.  
The reconstructed $D\ell$
samples contain, by design, the feed-down from all the higher mass
states (apart from decays of the type $\Bb\to\Ds X\ell\nub$~\cite{ref:DsKlnu}).
Kinematic restrictions are imposed to reduce the contribution of
backgrounds from semileptonic decays to final state hadronic
systems more massive than $\Dstar$ and from other sources of 
$D\ell$ combinations.  
Distributions from selected events are binned in the 3-dimensional
space described below. The electron and muon samples
are input into separate fits,  
in which isospin symmetry is assumed for the semileptonic decay rates.  
Semileptonic decays are produced via a spectator diagram in which the heavy 
quark decays independently; strong interaction corrections to this process 
conserve isospin. As a result, we constrain semileptonic decay rates for
$\Bm$ and $\Bzb$ to be equal, {\it e.g.}, 
$\Gamma(B^- \to D^0 l^- \nub) = \Gamma(\Bb^0 \to D^+ l^- \nub)$. 
This substantially reduces statistical uncertainties on the fitted parameters.
Systematic uncertainties associated with the modeling of the signal
and background processes, the detector response, and uncertainties on
input parameters are determined, along with their correlations between
the electron and muon samples.  The fitted results are then combined
using the full covariance matrix of statistical and systematic errors.
For both $\Bb\to D\ell\nub$ and $\Bb\to\Dstar\ell\nub$ decays,
the fitted branching fractions and
form factor parameters are used to determine the
products $\mathcal{G}(1)\Vcb$ and $\mathcal{F}(1)\Vcb$.  
These measurements, along with theoretical
input on the form factor normalizations $\mathcal{G}(1)$
and $\mathcal{F}(1)$ at the zero recoil point, allow
determinations of \Vcb.

The approach taken in this study has some similarity to that of
Ref.~\cite{ref:BaBarbclnutagged}, where the branching fractions for
$\Bb\to D\ell\nub$ and $\Bb\to\Dstar\ell\nub$ are measured
simultaneously.
However, Ref.~\cite{ref:BaBarbclnutagged} reconstructs semileptonic $B$ decays 
in events in which the second $B$ meson is fully
reconstructed.  
That approach allows the use of the missing mass
squared as a powerful discriminant.  This analysis
provides modest discrimination between the different semileptonic decays
on an event-by-event basis, but results in a much larger statistical
sample and enables the measurement of form factor parameters.

The remaining sections of this paper are organized as follows.  In
Sec.~\ref{sec:babar} we describe the \babar\ detector and the
samples of \babar\ data and simulated events used in the analysis.  The event
selection and the distributions that are input to
the global fit are discussed in Sec.~\ref{sec:selection}.
We give the parameterization of the form factors of 
$\Bb\to D^{(*)}\ell\nub$
decays and the modeling of semileptonic $\Bb$ decays to $D^{(*)}\pi$
and $D^{(*)}\pi\pi$ states in Sec.~\ref{sec:formfactors}.
The global fit strategy and results are given in
Sec.~\ref{sec:fit}, and the evaluation of systematic uncertainties
is detailed in  Sec.~\ref{sec:systematics}.  
Sec.~\ref{sec:physics} presents the determination of \Vcb\ from the 
fitted results.
The final section (\ref{sec:discussion}) discusses the results
and provides averages with previous \babar\ measurements.

\section{\boldmath The \babar\ detector and dataset}
\label{sec:babar}
The data used in this analysis were collected with the \babar\
detector at the \pep2 storage ring between 1999
and 2004.  
\pep2 is an asymmetric collider; 
the center-of-mass of the colliding $\epem$ moves with
velocity $\beta=0.49$ along the beam axis in the laboratory rest frame.
The data collected 
at energies near the peak of the
$\FourS$ resonance (on-peak) correspond to $207\invfb$ or 230 million $\BB$
decays.  Data collected just below $\BB$ threshold (off-peak),
corresponding to $21.5\invfb$, are used to subtract the
$\epem\to\qqbar$ ($q=u,d,s,c$) background under the $\FourS$
resonance.

The \babar~detector is described in detail elsewhere~\cite{ref:babar}.
It consists of a silicon vertex tracker (SVT), a drift chamber (DCH),
a detector of internally reflected Cherenkov light (DIRC), an
electromagnetic calorimeter (EMC) and an instrumented flux return
(IFR).  The SVT and DCH operate in an axial magnetic field of 1.5 T
and provide measurements of the positions and
momenta of charged particles, as well as of their ionization energy
loss ($dE/dx$).  Energy and shower shape measurements for photons
and electrons are provided by the EMC.  The DIRC measures the angle of
Cherenkov photons emitted by charged particles traversing the fused
silica radiator bars.  
Charged particles that traverse the EMC and showering hadrons are measured
in the IFR as they penetrate successive layers of the return yoke of the 
magnet.

Simulated events 
used in the analysis 
are generated using the EVTGEN~\cite{ref:evtgen} program, and the
generated particles are propagated through a model of the
\babar~detector with the GEANT4~\cite{ref:geant4} program and
reconstructed using the same algorithms used on \babar~data.
The form factor parameterization~\cite{ref:CLN}
used in the simulation for $\Bb\to\Dstar\ell\nub$ decays is based on 
Heavy Quark Effective Theory (HQET)~\cite{ref:Neubert},
while the ISGW2 model~\cite{ref:ISGW2} is used for 
$\Bb\to D\ell\nub$ and $\Bb\to D^{**}\ell\nub$ decays,
where $D^{**}$ is one of the four $P$-wave charm mesons as described 
in Sec.~\ref{subsec:DLcand}. 
These are subsequently reweighted to the forms given
in Sec.~\ref{sec:formfactors}.
For non-resonant $\Bb\to D^{(*)}\pi\ell\nub$ decays, 
the Goity-Roberts model~\cite{ref:GoityRoberts} is used.
In order to saturate the inclusive semileptonic $b\to c\ell\nub$
decay rate we include a contribution from $\Bb\to D^{(*)}\pi\pi\ell\nub$ 
decays; a variety of models are considered for this purpose.
The branching fractions for $\B$ and charm decays 
in the simulation are rescaled
to the
values in Ref.~\cite{ref:PDG2008}.  In addition, the 
momentum spectra for $D^0$ and $D^+$ from
$\B\to DX$ and $\Bb\to DX$ decays are adjusted to agree with the
corresponding measured spectra from
Ref.~\cite{ref:BaBarInclusiveBtoD}. 
This adjustment is done only for background processes. 

The simulation of the detector response provided by the GEANT4-based
program is further adjusted by comparing with \babar~data 
control samples.  In particular, the efficiency of charged track
reconstruction is modified by 1-2\%, depending on momenta and event 
multiplicity, based on studies of 
multi-hadron events and 1-versus-3
prong $\epem\to\taup\taum$ events.
The efficiencies and misidentification probabilities of the particle
identification (PID) algorithms used to select pions, kaons, electrons and
muons (see  Sec.~\ref{sec:selection}) are adjusted based on studies
of samples of $\epem\to\epem\gamma$ and $\epem\to\mup\mun\gamma$, 
and several samples reconstructed without particle identification:
1-versus-3 prong $\epem\to\taup\taum$ events, 
$\KS\to\pip\pim$, $\Dstarp\to\Dz\pip\to(\Km\pip)\pip$ 
and $\Lambda\to p\pim$.

\section{\boldmath Event selection}
\label{sec:selection}

\subsection{\boldmath Preselection of $D\ell$ candidates}
We select multi-hadron events by requiring 
at least three good-quality charged tracks,
a total reconstructed energy in the event exceeding 4.5~GeV,
the second normalized Fox-Wolfram moment~\cite{ref:FoxWolfram} $R_2 < 0.5$, 
and the distance between the interaction point and the primary vertex 
of the $B$ decay to be less than 0.5~cm (6.0~cm) in the direction
transverse (parallel) to the beam line.
In these events an identified electron or muon
candidate must be present, along with a candidate $D$ meson decay.
Candidate electrons are identified using a likelihood ratio based on
the shower shape in the EMC, $dE/dx$ in the tracking detectors, the
Cherenkov angle and the ratio of EMC energy to track momentum.  The
electron identification efficiency is $94\%$ within the acceptance of
the calorimeter, and the pion misidentification rate 
is
$0.1\%$.  Muon candidates are identified using a neural network that
takes input information from the tracking detectors, EMC, and IFR.
The muon identification efficiency rises with momentum to reach a
plateau of $70\%$ for laboratory momenta above $1.4\gev/c$, and the pion
misidentification rate 
is $3\%$.  

Kaon candidates
are required to satisfy particle identification criteria based
on the $dE/dx$ measured in the tracking detectors and the
Cherenkov angle measured in the DIRC.  Each kaon candidate is
combined with one or two charged tracks of opposite sign to form 
a $\Dz\to\Km\pip$ or $\Dp\to\Km\pip\pip$ candidate.  
Those combinations with invariant masses in 
the range
$1.840<m_{K\pi}<1.888\,\gev/c^2$ 
are considered as $\Dz$ 
candidates and those in the range $1.845<m_{K\pi\pi}<1.893\,\gev/c^2$
as $\Dp$ candidates, respectively.  Combinations
in the ``sideband'' mass regions
$1.816<m_{K\pi}<1.840\,\gev/c^2$ and $1.888<m_{K\pi}<1.912\,\gev/c^2$
($1.821<m_{K\pi\pi}<1.845\,\gev/c^2$ and $1.893<m_{K\pi\pi}<1.917\,\gev/c^2$)
are used to estimate the combinatorial background.

The charge of the kaon candidate
is required to have the same sign as that of the candidate lepton.
Each $D$--lepton combination in an event is
fitted to both $\Bb\to D\ell$ and $D\to\Km\pip(\pip)$ vertices
using the algorithm described in Ref.~\cite{ref:treefitter}.  The
fit probabilities are required to exceed 0.01 for the 
$\Bb\to \Dz\ell$ and $\Bb\to\Dp\ell$
vertices and 0.001 for the $\Dz$ and $\Dp$ decay vertices.  We require the
absolute value of the cosine of the angle between the $D\ell$ momentum 
vector and the thrust axis of the remaining particles in the event to be 
smaller than $0.92$ to further reduce
background, most of which comes from 
$\epem\to\qqbar$ ($q=u,d,s,c$) events.

The signal yields are determined by subtracting the estimated
combinatorial background from the number of $D$ candidates in the peak
region.  The combinatorial background is estimated using the number of
candidates in the $D$ mass sideband regions scaled by the ratio of the
widths of the signal and sideband regions.
This is
equivalent to assuming a linear dependence of the combinatorial
background on invariant mass.  The change in the yields is
negligible when using other assumptions for the background shape.
Candidates from $\epem\to\qqbar$ events are statistically removed from
the data sample by subtracting the distribution of candidates observed
in the data collected at energies below $\BB$ threshold (off-peak),
after scaling these data by the factor $r_{\cal L}=\left( {\cal
L}_\mathrm{on} {s_\mathrm{off}} \right) / \left( {\cal L}_\mathrm{off}
{s_\mathrm{on}} \right)$ to account for the difference in luminosity
and the dependence of the annihilation cross-section on energy.  
The selection criteria listed above were determined
using simulated $\BB$ events and off-peak data to
roughly maximize the
statistical significance of the $D\ell$ signal yields in
$\epem\to\BB$ events.  They have an overall
efficiency of $80\%$ ($76\%$) for $\Bb\to\Dz X\ell\nub$ 
($\Bb\to\Dp X\ell\nub$) decays 
with $p^*_\ell$, the lepton momentum magnitude in the center-of-mass (CM)
frame, in the range $0.8$--$2.8\gev/c$.

\begin{figure}[htbp]
\begin{center}
\scalebox{0.45}{\includegraphics{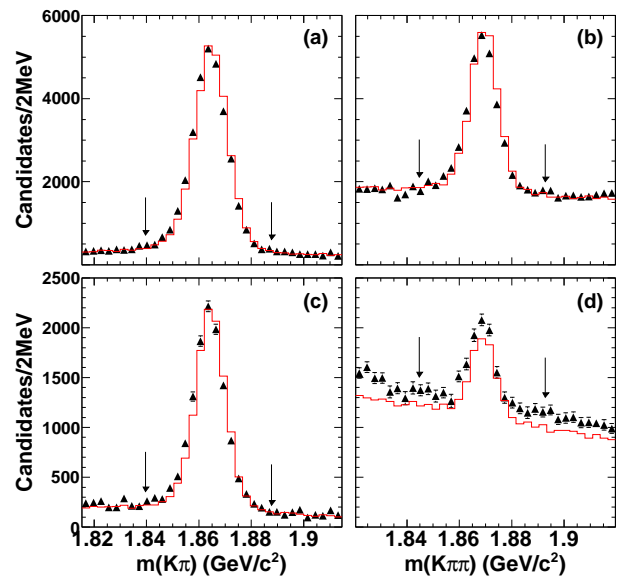}}
\caption{The invariant mass distributions (data points)
for selected candidates.
Scaled off-peak data have been subtracted
to remove contributions from $\epem\to\qqbar$ annihilation.
Plots (a,c) show $\Km\pip$ combinations and (b,d) show
$\Km\pip\pip$ combinations.  In each case the $D\ell$ candidates
are required to satisfy $-2.0<\cosBDl<1.1$.  
The further kinematic requirements are
$1.6<p^*_\ell<1.8\,\gev/c$, $1.6<p^*_D<2.0\,\gev/c$ for plots
(a,b) and $2.0<p^*_\ell<2.35\,\gev/c$, $0.8<p^*_D<1.2\,\gev/c$
for plots (c,d).
The histograms show the contribution from simulated $\BB$ events
 scaled to the data luminosity.
The arrows indicate the boundaries between signal and sideband regions.
}
\label{fig:Dmass}
\end{center}
\end{figure}

The invariant mass distributions for the $\Dz$ and $\Dp$
candidates, after off-peak subtraction, are shown
in Fig.~\ref{fig:Dmass} for two kinematic subsets representing
regions with good and poor signal-to-background ratios.
The small differences in peak position and
combinatorial background level have negligible impact on the analysis 
due to the sideband subtraction described above and the wide signal window.

The $\Dz\ell$ and $\Dp\ell$ candidates are binned
in three kinematic variables:
\begin{itemize} 
\item $p^*_D$, the $D$ momentum in the CM frame;
\item $p^*_\ell$, the lepton momentum in the CM frame;
\item 
$
\cosBDl \equiv \left(2 E^*_B E^*_{D\ell} - m_B^2 - m_{D\ell}^2 \right)
\left/ \left(2 p^*_B p^*_{D\ell}\right)\right. ,
$
the cosine of the angle between the $\Bb$ and $D\ell$ momentum
vectors in the CM frame
under the assumption that the $\Bb$ decayed to $D\ell\nub$.
If the $D\ell$ pair is not from a $\Bb\to D\ell\nub$ decay, 
$|\cosBDl|$ can exceed unity.
The $B$ energy and momentum are not measured
event-by-event; they are calculated from the CM energy determined by
the \pep2 beams as $E^*_B = \sqrt{s}/2$ and $p^*_B =
\sqrt{{E^*_B}^2-m_B^2}$, where $m_B$ is the $B^0$ meson mass.
The energy, momentum and invariant mass
corresponding to the sum of the $D$ and lepton four vectors in the CM
frame are denoted $E^*_{D\ell}$, $p^*_{D\ell}$ and $m_{D\ell}$, 
respectively.  
\end{itemize} 
The binning in these three variables 
is discussed in Sec.~\ref{subsec:KinematicCut}.

\subsection{\boldmath Sources of $D\ell$ candidates}\label{subsec:DLcand}
There are several sources of $D\ell$ candidates that
survive the $D$-mass sideband and off-peak subtractions.  In both
the $\Dz$ and $\Dp$ samples we group them as follows ($\Bb$ 
represents both $\Bm$ and $\Bzb$):
\begin{enumerate}[(i)]
\item $\Bb\to D\ell\nub$ \label{it:BDlnu}
\item $\Bb\to \Dstar\ell\nub$ \label{it:BDslnu}
\item $\Bb\to D^{(*)}(n\pi)\ell\nub$, which includes\label{it:BDsslnu}
\begin{itemize}
\item The P-wave $D^{**}$ charm mesons.
In the framework of HQET, the P-wave charm mesons are categorized by the 
angular momentum of the light constituent, $j_{\ell}$, namely 
$j_{\ell}^P = 1/2^-$ doublet $D_0^*$ and $D^\prime_1$ and
$j_{\ell}^P = 3/2^-$ doublet $D_1$ and $D_2^*$~\cite{ref:LLSW}. 
\item Non-resonant $\Bb\to D^{(*)}\pi\ell\nub$.
\item Decays of the type $\Bb\to D^{(*)}\pi\pi\ell\nub$; the modeling of 
these is discussed in Sec.~\ref{subsec:Dpipilnu}.
\end{itemize}
\item Background from $\BB$ events in which the lepton and $D$ candidates
do not arise from a single semileptonic $\Bb$ decay.  These include (in order
of importance)\label{it:BBbkg}
\begin{itemize}
\item Direct leptons from $\Bb\to X\ell\nub$ decays combined with
a $D$ from the decay of the other $\B$ meson in the
event.  Roughly one third of this background comes from events in which
$\BzBzb$ mixing results in the decay of two $\Bzb$ mesons.  
Most of the remaining contribution comes from CKM-suppressed $\B\to DX$ 
transitions.  
\label{it:uncorrDL}
\item Uncorrelated cascade decays. 
In this case the lepton mostly
comes from the decay of an anti-charm meson 
produced in the $\B$ 
decay and the $D$ arises from the decay of the other $\Bb$ 
meson in the event.  
\label{it:uncorrCL}
\item Correlated cascade decays, in which the lepton and $D$
candidates come from the same parent $\Bb$ meson.  These are mainly
$\Bb\to D\Db (X)$ and $\Bb\to D(X)\tau\nu$ decays, with the lepton coming from
the decay of an anti-charm meson or tau.\label{it:corrCL}
\item Mis-identified lepton background.  The probability of a hadron
being misidentified as a lepton is negligible
for electrons but not for muons.\label{it:fake}
\end{itemize}
\end{enumerate}
As mentioned previously, 
the same decay 
widths are imposed for the semileptonic transitions of $\Bzb$ and $\Bm$.  
For
the background processes (source~\ref{it:BBbkg})
no such requirement is imposed.

\subsection{\boldmath Kinematic restrictions}\label{subsec:KinematicCut}
Despite the use of the best available information for calculating
the background and $\Bb\to D^{(*)}(n\pi)\ell\nub$ distributions, these
components suffer from significant uncertainties.  We therefore
restrict the kinematic range of the variables used in the fit
to reduce the impact of these uncertainties while preserving
sensitivity to the $\Bb\to D\ell\nub$ and $\Bb\to\Dstar\ell\nub$
branching fractions and form factor parameters.  We require
$-2<\cosBDl<1.1$ and place restrictions on $p^*_D$ and
$p^*_\ell$, rejecting regions where the signal decays 
are not dominant.
This results in the ranges $1.2\,\gev/c<p^*_\ell<2.35\,\gev/c$ and
$0.8\,\gev/c<p^*_D<2.25\,\gev/c$.  
The yield within this region is $4.79\times 10^5$ ($2.95\times 10^5$) 
candidates in the
$\Dz\ell$ ($\Dp\ell$) sample
with a statistical uncertainty of $0.26\%$ ($0.66\%$).
 
\begin{table}[!htb]
\caption{Definition of bins used for kinematic variables.}
\begin{tabular}{lcc} \hline\hline
Quantity & \# bins & Bin edges \\ \hline
$\cosBDl$ & 3 & $-2.0$, $-1.0$, $0.0$, $1.1$  \\
$p^*_\ell\ (\gev/c)$ 
& 10 & $1.2$,$1.3$,$1.4$,$1.5$,$1.6$,$1.7$,$1.8$,$1.9$,$2.0$,$2.1$,$2.35$ \\
$p^*_D\ (\gev/c)$
& 8 & $0.8$,$1.1$,$1.35$,$1.5$,$1.65$,$1.8$,$1.95$,$2.1$,$2.25$ \\
\hline\hline
\end{tabular}
\label{tab:binning}
\end{table}
The data are binned finely enough to have good sensitivity to the fit
parameters while maintaining adequate statistics per bin.  
Table~\ref{tab:binning} gives the binning used in the fit.
We avoid setting a bin
edge at $\cosBDl=1$ to reduce our sensitivity to the modeling
of the resolution in this variable, since the $\Bb\to D\ell\nub$
decay distribution has a sharp cut-off at this point.  

\begin{figure}[htbp]
\begin{center}
\scalebox{0.45}{\includegraphics{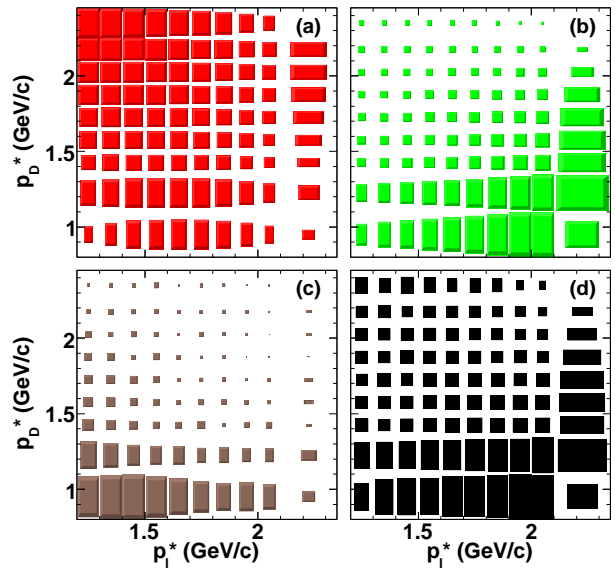}}
\caption{(Color online) Distribution of $p^*_D$ vs.~$p^*_\ell$
for $\Dz e$ candidates after sideband subtraction.
The shaded boxes have area proportional to the number of entries.  
The plots show simulated candidates for
(a) $\Bb\to D e\nub$,
(b) $\Bb\to D^* e\nub$ and (c) other (sources~\ref{it:BDsslnu}
and \ref{it:BBbkg} combined), and for data after off-peak subtraction (d).
The binning given in Table~\ref{tab:binning} is used 
and only candidates that satisfy $0.0<\cosBDl<1.1$ are plotted.
}
\label{fig:3dplot}
\end{center}
\end{figure}
\begin{figure*}[htbp]
\begin{center}
\scalebox{0.5}{\includegraphics{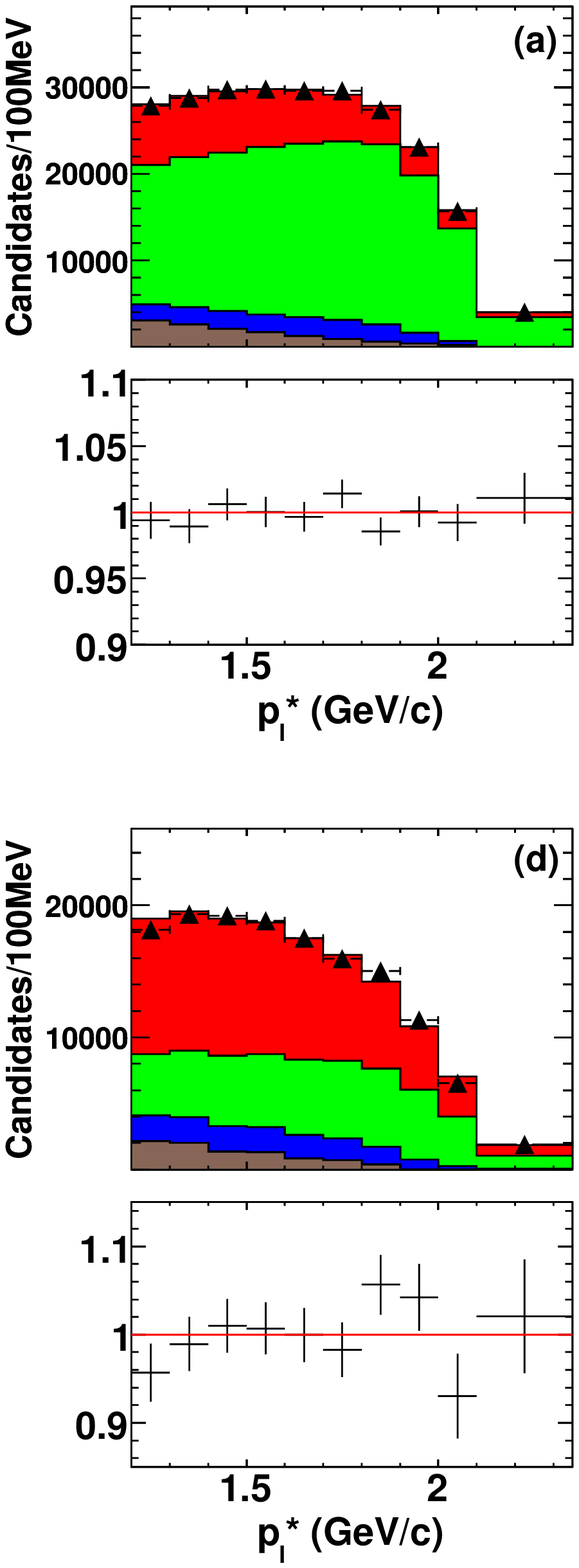}}
\scalebox{0.5}{\includegraphics{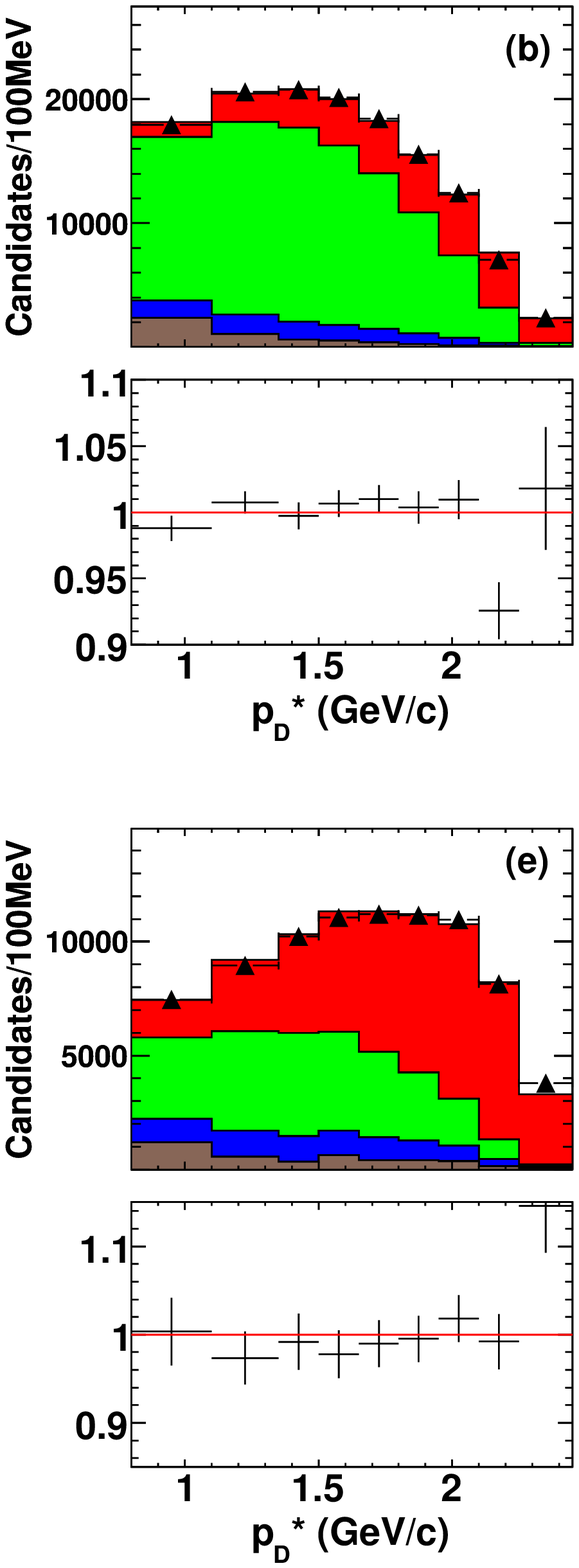}}
\scalebox{0.5}{\includegraphics{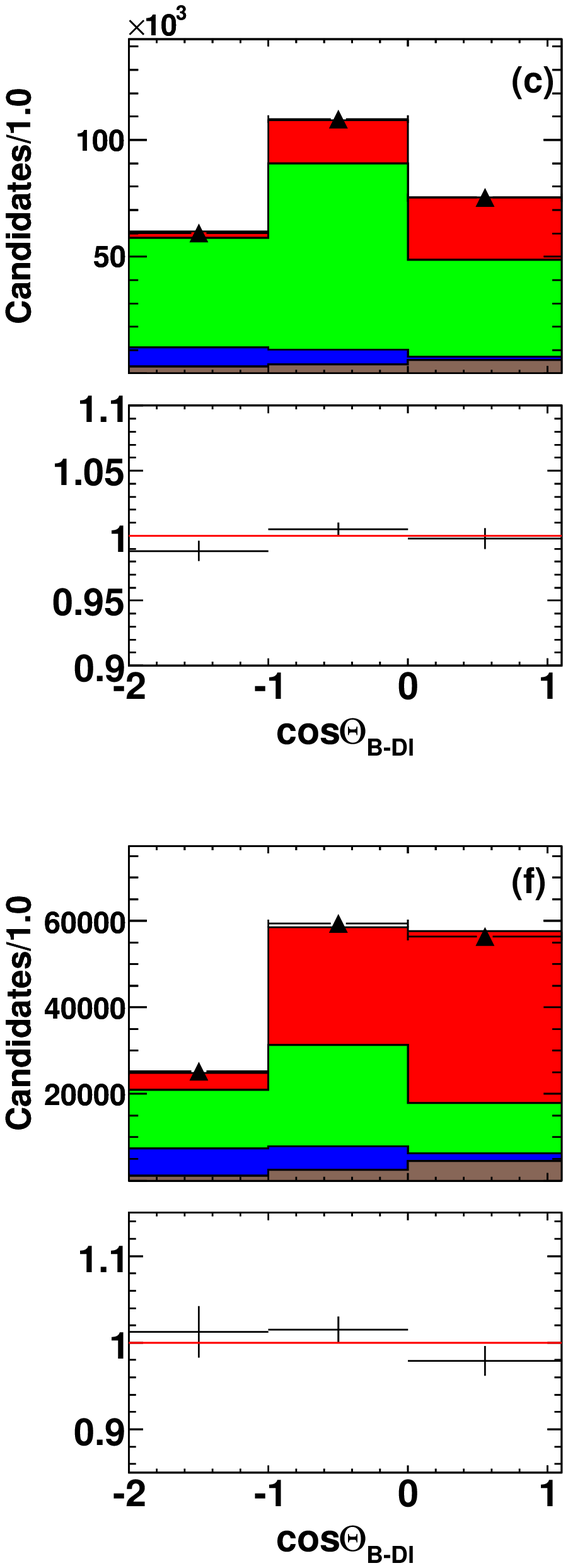}}
\caption{(Color online) Projections onto individual kinematic variables of the
data after off-peak subtraction and the results of the fit : 
(a,d) lepton and (b,e) $D$ momentum in the CM frame,
and (c,f) $\cos\theta_{B-D\ell}$. 
The points show data for accepted $\Dz e$ (a,b,c) 
and $\Dp e$ (d,e,f) candidates, and the histograms show the individual 
fit components (from top to bottom): $\Bb\to D e\nub$, 
$\Bb\to \Dstar e\nub$, 
$\Bb\to D^{(*)}(n\pi) e\nub$ 
and other $\BB$ background. 
\hspace{0.02cm}
The ratio of data to the sum of the fitted yields is shown 
below each plot.
}
\label{fig:1d}
\end{center}
\end{figure*}
\begin{figure*}[htbp]
\begin{center}
\scalebox{0.5}{\includegraphics{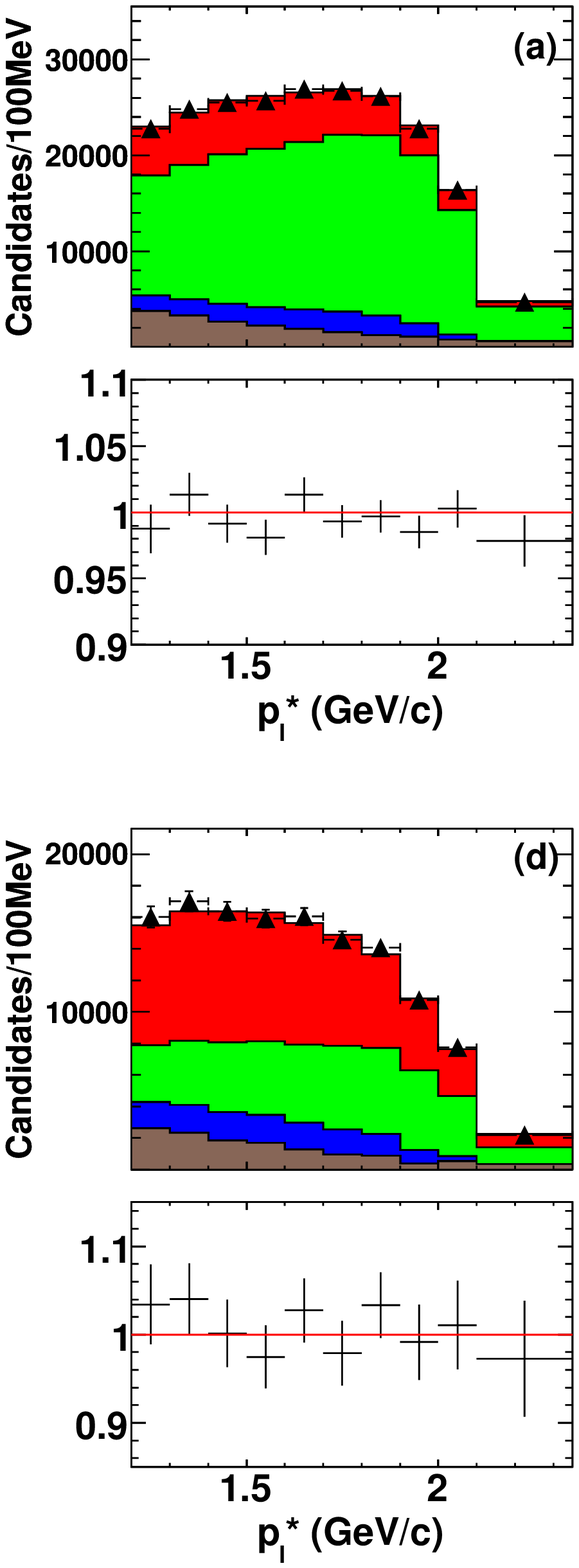}}
\scalebox{0.5}{\includegraphics{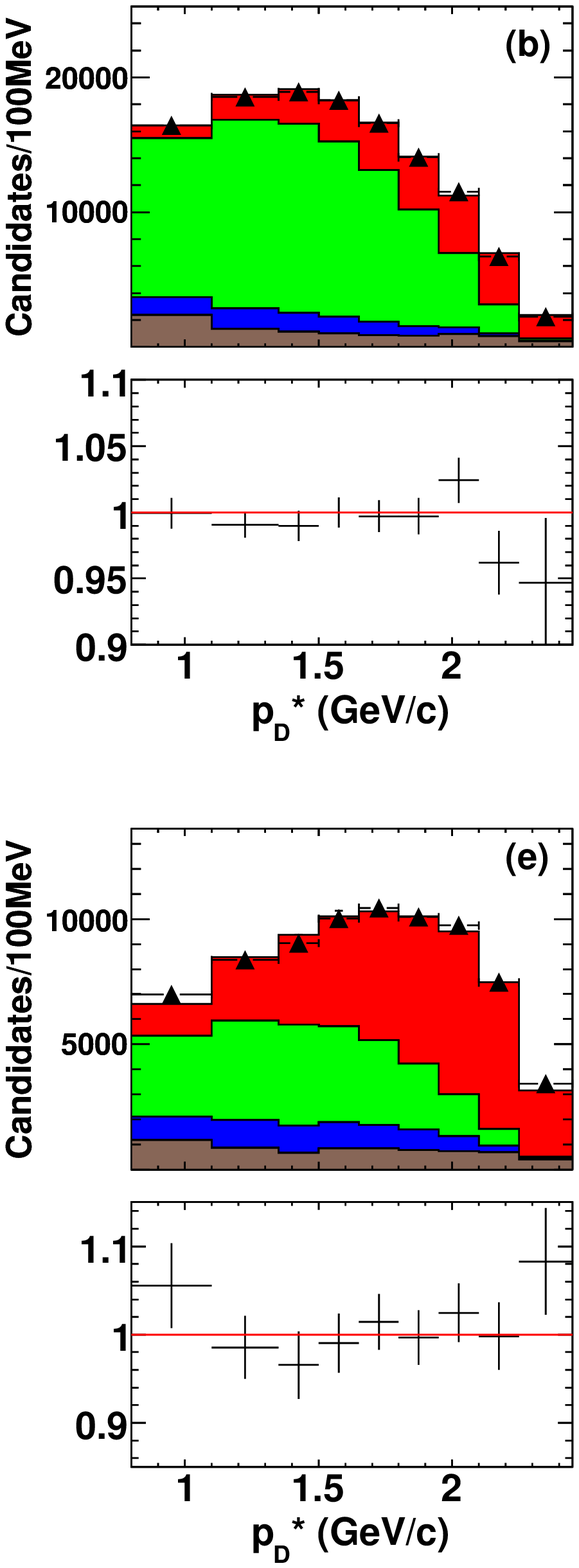}}
\scalebox{0.5}{\includegraphics{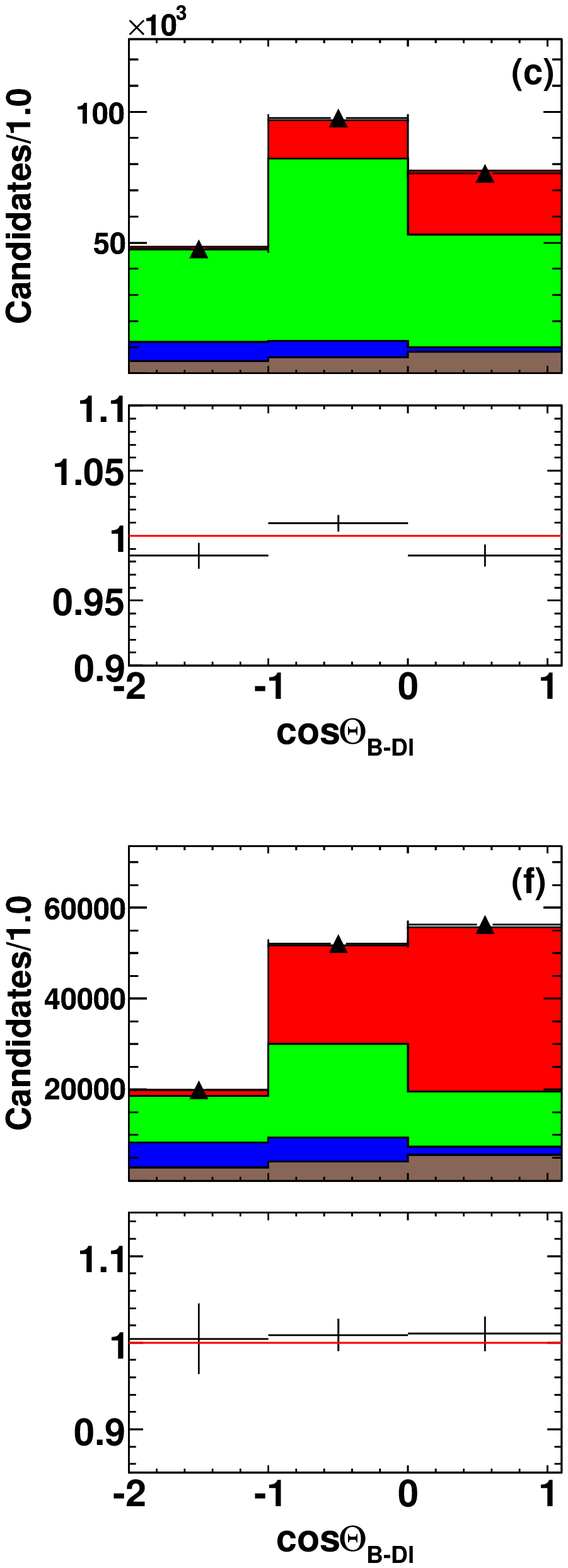}}
\caption{(Color online) Projections onto individual kinematic variables of the
data after off-peak subtraction and the results of the fit : 
(a,d) lepton and (b,e) $D$ momentum in the CM frame,
and (c,f) $\cos\theta_{B-D\ell}$. 
The points show data for accepted $\Dz\mu$ (a,b,c) 
and $\Dp\mu$ (d,e,f) candidates, and the histograms show the individual 
fit components (from top to bottom): $\Bb\to D \mu\nub$, 
$\Bb\to \Dstar \mu\nub$, 
$\Bb\to D^{(*)}(n\pi) \mu\nub$ 
and other $\BB$ background.
\hspace{0.02cm}
The ratio of data to the sum of the fitted yields is shown 
below each plot.
}
\label{fig:1dMu}
\end{center}
\end{figure*}
Two-dimensional projections of the signal, background and
data distributions for the $\Dz e$ sample
are shown in Fig.~\ref{fig:3dplot} to illustrate
the separation power in these variables.  
The distributions for the $\Dz \mu$ sample (not shown) 
are similar. 
The one-dimensional projections of the $De$ and $D\mu$ samples are shown in
Figs.~\ref{fig:1d} and \ref{fig:1dMu}.
The difference in the size of the $\Bb\to\Dstar\ell\nub$ components in
$D^0\ell$ and $D^+\ell$ distributions is due to the fact that $D^{*0}$
does not decay to $D^+$.

\section{\boldmath Modeling of semileptonic $B$ decays}
\label{sec:formfactors}
In our fully simulated event samples 
$\Bb\to D\ell\nub$ and $\Bb\to D^{**} \ell \nub$ decays were
generated using the ISGW2 model~\cite{ref:ISGW2}.
For $\Bb\to\Dstar\ell\nub$ decays, an HQET model was used with a 
linear form factor parameterization. 
We re-weight all these decays using the
formulae given in the following subsections.
The histograms in Figs.~\ref{fig:1d} and \ref{fig:1dMu} are re-weighted.

\subsection{$\Bb\to D\ell\nub$ decays}
\label{subsec:Dlnu}
The differential decay rate is given by~\cite{ref:Neubert}
\begin{eqnarray}
  \frac{d\Gamma(\Bb\to D\ell\nub)}{dw} 
  = \frac{G_F^2 |V_{cb}|^2 m_B^5}{48\pi^3}
    r^3 (w^2 - 1)^{3/2}   \hspace{0.5cm} \nonumber \\
    \times [(1+r) h_+(w) - (1-r)h_-(w)]^2,
\end{eqnarray}
where $G_F$ is the Fermi constant, 
$h_+(w)$ and $h_-(w)$ are the form factors, $r\equiv m_D/m_B$ is the mass ratio
and $m_B$ and $m_D$ are the $B$ and $D$ meson masses, respectively.
The velocity transfer $w$ is defined as
\begin{equation}
   w \equiv v_B \cdot v_{D},
\end{equation}
where $v_B$  and $v_D$ are the 4-velocities of the $B$ and $D$ mesons,
respectively.  In the $B$ rest frame $w$ corresponds to the Lorentz boost
of the $D$ meson.
In the HQET model, the form factors are given by~\cite{ref:CLN}
\begin{equation}
\begin{array}{l}
  h_{+}(w) 
  = \mathcal{G}(1) \times  \\
    \hspace{0.3cm}
       [1 - 8 \rho_{D}^2 z + (51 \rho_{D}^2 - 10) z^2
            - (252\rho_{D}^2 - 84) z^3] 
\end{array}
\end{equation}
and
\begin{equation}
  h_{-}(w) = 0,
\end{equation}
where
$z = (\sqrt{w + 1} - \sqrt{2})/(\sqrt{w + 1} + \sqrt{2})$ and
$\rho_{D}^2$ and $\mathcal{G}(1)$ are, respectively, the
form factor slope and normalization at $w=1$.

The above formulae neglect the lepton mass $m_{\ell}$. 
Muon mass effects need to be included to achieve precision at the
few percent level on the form factor parameters.  Allowing for
non-zero lepton mass introduces additional terms in the phase space and 
form factor expressions~\cite{ref:KS1990MuonMass}
that can be included
by multiplying the decay rate formula by the following factor:
\begin{equation}
  W_D = \left( 1 - \frac{1}{1+r^2 - 2rw}
               \frac{m_{\ell}^2}{m_B^2}
     \right)^2
     \left[1 + \mathcal{K}_D(w)\frac{m_{\ell}^2}{m_B^2}\right]
\end{equation}
where
\begin{equation}
  \mathcal{K}_D(w) \equiv
  \left[
  1 + 
  3 \left( \frac{1-r}{1+r} \right)^2
    \left( \frac{w+1}{w-1} \right)
  \right] 
  \frac{1}{2(1+r^2 - 2rw)}.
\end{equation}

\subsection{$\Bb\to\Dstar\ell\nub$ decays}
\begin{figure}[htp]
\begin{center}
\scalebox{0.5}{\includegraphics{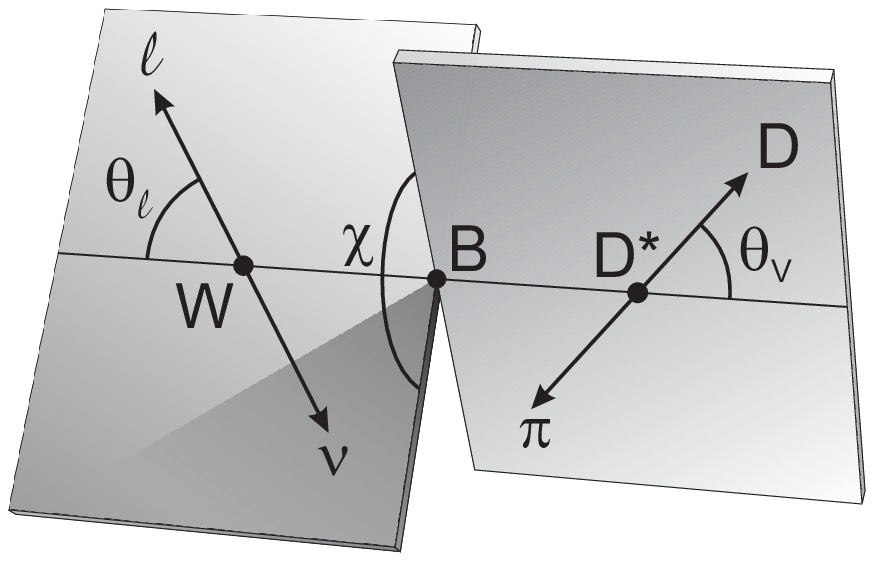}}
\caption{\label{DecayGeom} Definition of the three angles
$\theta_\ell$, $\theta_V$ and $\chi$ for the decay $\Bb\to\Dstar\ell\nub$.}
\end{center}
\end{figure}

We need three additional kinematic variables to describe this decay. 
A common choice is
$\theta_{\ell}$, $\theta_{V}$ and $\chi$, shown in
Fig.~\ref{DecayGeom}, and defined as
\begin{itemize}
\item $\theta_{\ell}$ : the angle between 
the lepton and the direction opposite the $B$ meson in the $W$ rest frame.
\item $\theta_V$ : the angle between 
the $D$ meson and the direction opposite the $B$ meson in the $D^*$ rest frame.
\item $\chi$ : the azimuthal angle between the planes formed by the
$W$-$\ell$ and $D^*-D$ systems in the $B$ rest frame.
\end{itemize}
\label{subsec:Dslnu}
The differential decay rate is given by~\cite{ref:Neubert}
\begin{eqnarray}
  \frac{d\Gamma(\Bb\to\Dstar\ell\nub)}
      {dw\,d\mathrm{cos}\theta_{V}\,d\mathrm{cos}\theta_{\ell}\,d\chi}
                                                \hspace{5.0cm}\nonumber\\
    = \frac{3G^{2}_{F}}{4(4\pi)^{4}}
         |V_{cb}|^{2} m_B m_{D^*}^2 
	 \sqrt{w^2 - 1} 
	 (1 + r^{*2} - 2r^*w) \times \hspace{0.0cm}\nonumber\\
    \lbrack(1-\mathrm{cos}\theta_{\ell})^{2}\mathrm{sin}^{2}\theta_{V}
                                 |H_{+}(w)|^{2}  \hspace{3.1cm}\nonumber\\
    +(1+\mathrm{cos}\theta_{\ell})^{2}\mathrm{sin}^{2}\theta_{V}
                                 |H_{-}(w)|^{2}  \hspace{2.75cm}\nonumber\\
    +4\mathrm{sin}^{2}\theta_{\ell}\mathrm{cos}^{2}\theta_{V}
                                 |H_{0}(w)|^{2} \hspace{3.5cm}\nonumber\\
    -4\mathrm{sin}\theta_{\ell}(1-\mathrm{cos}\theta_{\ell})
        \mathrm{sin}\theta_{V}\mathrm{cos}\theta_{V}\mathrm{cos}\chi
                               H_{+}(w)H_{0}(w)  \hspace{0.0cm}\nonumber\\
    +4\mathrm{sin}\theta_{\ell}(1+\mathrm{cos}\theta_{\ell})
        \mathrm{sin}\theta_{V}\mathrm{cos}\theta_{V}\mathrm{cos}\chi
                               H_{-}(w)H_{0}(w)  \hspace{0.0cm}\nonumber\\
    -2\mathrm{sin}^{2}\theta_{\ell}\mathrm{sin}^{2}\theta_{V}\mathrm{cos}2\chi
        H_{+}(w)H_{-}(w)\rbrack,             \hspace{1.75cm}
\end{eqnarray}
where $H_i(w)$ are form factors, $r^*=m_{D^*}/m_B$
and $m_{D^*}$ is the $D^*$ meson mass.
The $H_i(w)$ are usually written in terms of one form factor 
$h_{A_1}(w)$ and two form factor ratios, $R_1(w)$ and
$R_2(w)$, as follows:
\begin{equation}
  H_i = -m_B
        \frac{R^* (1 - r^{*2}) (w + 1)}{2 \sqrt{1 + r^{*2} - 2r^*w}}
	h_{A_1}(w) \widetilde{H}_i(w),
\end{equation}
where $R^* = (2 \sqrt{m_B m_{D^*}})/(m_B + m_{D^*})$ and 
\begin{eqnarray}
  \widetilde{H}_{\pm}(w)
  & = & \frac{\sqrt{1 + r^{*2} - 2r^*w}}{1 - r^*}
    \left( 1 \mp \sqrt{\frac{w-1}{w+1}} R_{1}(w) \right),\nonumber \\
  \widetilde{H}_{0}(w)
  & = & 1 + \frac{w - 1}{1 - r^*} \left( 1 - R_{2}(w) \right) .
\end{eqnarray}
The form factor ratios have a modest 
dependence on $w$, estimated~\cite{ref:CLN} as
\begin{eqnarray}
  R_1(w) & = & R_1 - 0.12(w-1) + 0.05(w-1)^2, \nonumber \\
  R_2(w) & = &R_2 + 0.11(w-1) - 0.06(w-1)^2.
\end{eqnarray}
The form used for $h_{A_1}(w)$ is~\cite{ref:CLN}
\begin{eqnarray}
  h_{A_1}(w) 
  = \mathcal{F}(1) \times \hspace{5cm}\nonumber \\
      \left[ 1 - 8 \rho_{D^*}^2 z + (53 \rho_{D^*}^2 - 15) z^2
            - (231\rho_{D^*}^2 - 91) z^3 \right],
\end{eqnarray}
where $\rho_{\Dstar}^2$ and $\mathcal{F}(1)$ are, respectively, the
form factor slope and normalization at $w=1$.

Non-zero lepton mass is accounted for by multiplying 
the decay rate formula by the factor
\begin{equation}
  W_{D^*} = \left( 1 - \frac{1}{1+r^{*2} - 2r^*w}
               \frac{m_{\ell}^2}{m_B^2}
     \right)^2
     \left[1 + \mathcal{K}_{D^*}(w)\frac{m_{\ell}^2}{m_B^2}\right],
\end{equation}
where
\begin{equation}
  \mathcal{K}_{D^*}(w) \equiv
  \left[
  1 + 
  \frac{3}{2} 
  \frac{\tilde{H}_t^2}{\tilde{H}_+^2 + \tilde{H}_-^2 +\tilde{H}_0^2}
  \right] 
  \frac{1}{2(1+r^{*2} - 2r^*w)} .
\end{equation}
Here, $\tilde{H}_t$ is expressed, using another form factor ratio $R_3(w)$, by
\begin{equation}
  \tilde{H}_t (w)
  = \frac{\sqrt{w^2 - 1}}{1-r^*}
    \left(
      1 + \frac{r^*-w}{w+1}R_3 - \frac{1+r^{*2}-2r^*w}{r^*(w+1)}R_2
    \right) .
\end{equation}
We take $R_3(w) = 1$; this approximation
has a negligible impact on our fit results.

\subsection{$\Bb\to D^{(*)}\pi\ell\nub$ decays}
\label{subsec:Dpilnu}
The four P-wave $D^{**}$ states
have been measured
in semileptonic decays~\cite{ref:HFAG,ref:BtoNarrowDsslnu,ref:BtoDsslnu}.
The decays $\Bb\to D^{**}\ell\nub$ are modeled following an HQET-inspired form
factor parameterization given in Ref.~\cite{ref:LLSW}.  Detailed
formulae are given in Appendix~\ref{sec:LLSWmodel}.
We use the
approximation B$_1$ of this model for our main fit and use the
approximation B$_2$ to evaluate the uncertainty due to the
approximation.  The slope of the form factors versus $w$ is parameterized
by $\hat{\tau}'$, which we set to $-1.5$ and vary between $-1.0$ and
$-2.0$ to study systematic uncertainties (Table~\ref{tab:inputParms}).

To parameterize the $\Bb\to D^{(*)}\pi\ell\nub$ decay branching fractions we
define five branching fraction ratios:
\begin{equation}
  f_{D_2^*/D_1} 
  \equiv \frac{\mathcal{B}(B^-\to{D}_2^{*0}\ell\nub)}
              {\mathcal{B}(B^-\to{D}_1^0\ell\nub)},
     \label{eq_fBpDD12} 
\end{equation}
\begin{equation}
  f_{D\pi/D_0^*} \equiv \frac{\mathcal{B}^{NR}(B^-\to D^{+}\pi^-\ell\nub)}
{\mathcal{B}(B^-\to{D}_0^{*0}\ell\nub)}, 
\end{equation}
\begin{equation}
  f_{D^*\pi/D_1'} 
  \equiv \frac{\mathcal{B}^{NR}(B^- \to D^{*+} \pi^- \ell \nub)}
	      {\mathcal{B}(B^- \to D_1'^{0} \ell \nub)}, 
\end{equation}
\begin{equation}
  f_{D_0^* D\pi/D_1 D_2^*}
   \equiv
   \frac{\mathcal{B}(B^-\to D_0^{*0}\ell\nub)
          + \mathcal{B}^{NR}(\Bm\to D\pi\ell\nub)}
        {\mathcal{B}(B^-\to{D}_1^0\ell\nub) 
          + \mathcal{B}(B^-\to{D}_2^{*0}\ell\nub)}, 
\end{equation}
\begin{equation}
  f_{D_1' D^*\pi/D_1 D_2^*}
  \equiv
   \frac{\mathcal{B}(B^- \to {D}_1'^{0} \ell \nub) 
          + \mathcal{B}^{NR}(\Bm\to {D}^* \pi \ell \nub)}
        {\mathcal{B}(B^-\to{D}_1^0\ell\nub) 
          + \mathcal{B}(B^-\to{D}_2^{*0}\ell\nub)}, 
\end{equation}
where NR stands for ``non-resonant'' decays, which are assumed to be
isospin-invariant.  The quantity 
$f_{D_2^*/D_1}$ is the ratio between two narrow states,
$f_{D\pi/D_0^*}$ ($f_{D^*\pi/D_1'}$) is between two broad states 
decaying to $D\pi$ ($D^*\pi$)
and the other two ratios are between broad and narrow states.
With these definitions the branching fractions 
for individual modes can be related to 
the total branching fraction
$\mathcal{B}(B^- \to D^{(*)} \pi \ell \nub)\equiv 
\mathcal{B}(B^- \to D \pi \ell \nub)+
\mathcal{B}(B^- \to D^* \pi \ell \nub)$.
We combine a new measurement~\cite{ref:BaBarbclnutagged} with the
world average~\cite{ref:PDG2008}
to determine 
the value given in Table~\ref{tab:inputParms}.

To estimate the branching fraction ratios, we average several 
measurements~\cite{ref:HFAG,ref:BtoNarrowDsslnu,ref:BtoDsslnu,ref:BELLE_BtoDsslnu} to find
\begin{eqnarray}
  \mathcal{B}(B^-\to D_1^0\ell\nub) &=& 0.0042 \pm 0.0004, \nonumber \\
  \mathcal{B}(B^-\to D_2^{*0}\ell\nub) &=& 0.0031 \pm 0.0005, \\
  \mathcal{B}(B^-\to D_1'^0\ell\nub) &=& 0.0022 \pm 0.0014, \nonumber \\
  \mathcal{B}(B^-\to D_0^{*0}\ell\nub) &=& 0.0048 \pm 0.0008.\nonumber
           \label{DDstarlnuBFNew}
\end{eqnarray}
The sum of the $D^{**}$ branching fractions saturates 
$\mathcal{B}(B^-\to D^{(*)} \pi \ell \nub)$ 
which implies that the non-resonant branching fractions are small. We use
\begin{eqnarray}
  \mathcal{B}^{NR}(B^- \to  D \pi\ell \nub)
    &=& 0.0015 \pm 0.0015\nonumber \\
  \mathcal{B}^{NR}(B^- \to  D^* \pi \ell \nub)
    &=& 0.00045 \pm 0.00045.
  \label{NRDpilnuBFNew}
\end{eqnarray}
From these numbers the branching fraction ratios are calculated
and listed in Table~\ref{tab:inputParms}.
These quantities are taken as independent when evaluating
systematic uncertainties.

\subsection{$\Bb\to D^{(*)}\pi\pi\ell\nub$ decays}
\label{subsec:Dpipilnu}
Recent measurements~\cite{ref:PDG2008,ref:BaBarbclnutagged} indicate
that the inclusive $\Bb\to X_c\ell\nub$ branching 
fraction is not saturated by the sum of the
$\Bb\to D\ell\nub$, $\Bb\to \Dstar\ell\nub$ and $\Bb\to D^{(*)}\pi\ell\nub$ 
branching fractions. 
In order to fill the gap, we include $\Bb\to D^{(*)}\pi\pi\ell\nub$ decays 
in our fit. 
We assume the branching fraction of these decays,
given in Table~\ref{tab:inputParms},
is equal to this
missing contribution to the inclusive branching fraction~\cite{ref:PDG2008}.

The $\Bb\to D^{(*)}\pi\pi\ell\nub$ decays are modeled as a combination
of four resonances : pseudo-scalar ($X_c$) and 
vector ($X_c^*$) states just above
$D^*\pi\pi$ threshold, and a heavier pair of pseudo-scalar ($Y_c$) 
and vector ($Y_c^*$) states just above $D^*\rho$ threshold, as listed 
in Table~\ref{XcStates}.  
Each state is assumed to be produced with equal rate
in semileptonic $B$ decays and each is assumed to decay with equal branching
fraction to $D\pi\pi$ and $D^*\pi\pi$, conserving isospin.
These assumptions are varied in assessing systematic uncertainties.

\section{\boldmath Global fit}
\label{sec:fit}

The binned distributions of $\Dz\ell$ and $\Dp\ell$ candidates in the
variables $p_\ell^*$, $p_D^*$ and $\cosBDl$ are fitted with the sum
of distributions for the signal and background 
sources listed in  Sec.~\ref{subsec:DLcand}.
The expected shape of the individual components is based on simulation,
and the fit adjusts the normalization of each component to minimize
the global chi-squared:
\begin{equation} \label{eq:chi2}
 \chi^2(\vec{\alpha}) 
 = \sum_{\mathrm{bin\ } i} 
 \frac{\left(N_i^\mathrm{on} - r_{\cal L} N_i^\mathrm{off} -
  \sum_{j} r_{j} C_j M_{ij} \right)^2}
 {(\sigma_i^\mathrm{on})^2+ r_{\cal L}^2 (\sigma_i^\mathrm{off})^2 +
  \sum_{j} r_{j}^2 C_j^2 (\sigma_{ij}^\mathrm{MC})^2 },
\end{equation}
where the index $i$ sums over bins of the $\Dz\ell$ and $\Dp\ell$
distributions and $j$ sums over individual simulated components.  The
coefficients $C_j$ depend on $\vec{\alpha}$, the set of free parameters
determined by minimizing $\chi^2$.  For example, for the
$\Bm\to\Dz\ell\nub$ component the coefficient $C_j$ is given by the
ratio of the fitted $\mathcal{B}(\Bm\to\Dz\ell\nub)$ branching
fraction to the value used in generating the corresponding
distribution.  The number of candidates in the data collected
on (below) the $\FourS$ peak in bin $i$ is denoted $N_i^\mathrm{on}$
($N_i^\mathrm{off}$) and $M_{ij}$ is the number of simulated events in
bin $i$ from source $j$.  
The $M_{ij}$ may depend on $\vec{\alpha}$ as explained below. 
The statistical 
uncertainties, after $D$ mass
sideband subtraction, are given by the $\sigma_i$ for the data
and the $\sigma_{ij}$ for the different Monte Carlo samples.  
The factor $r_{j}$
is the ratio of the on-peak luminosity to the effective luminosity of
the appropriate Monte Carlo sample.  Only those bins in which the
number of entries expected from the simulation exceeds $10$ are used
in the $\chi^2$ sum.

For the $\Bb\to D\ell\nub$ and $\Bb\to\Dstar\ell\nub$ signal components
we fit for both the branching fractions and for form-factor parameters.  
To facilitate this, we split these components into sub-components, one
corresponding to each unique combination of the parameters $\vec{\alpha}$ 
in the expression for the decay rate.
In terms of the notation
used in Eq.~\ref{eq:chi2}, we set
\begin{equation}
 C_j M_{ij}
 = \sum_k C_{j}^{(k)} M_{ij}^{(k)},
\end{equation}
where the index $k$ runs over the sub-components.
For example, the form factor in
$\Bb\to D\ell\nub$ decays is of the form $\mathcal{G}(z,\rho_D^2)=A(z)-\rho_D^2
B(z)$, where $\rho_D^2$, the slope of the form factor, is a parameter in
the fit and $z$ is a kinematic variable.  
The decay rate, which depends on the square of $\mathcal{G}$, 
has terms proportional to $1$, $\rho_D^2$ and
$(\rho_D^2)^2$, thus requiring three sub-components 
with coefficients $C_{j}^{(1)}$ to $C_{j}^{(3)}$.
The calculation of the variance for the $\Bb\to D\ell\nub$ component
involves the fourth power of $\mathcal{G}$ and thus
requires five sub-components.  For the
$\Bb\to\Dstar\ell\nub$ decay we use 18 sub-components to allow the
fitting of the form factor parameters $R_1$, $R_2$ and
$\rho^2_{\Dstar}$ and 75 sub-components to calculate the associated
variance.  By breaking the components up in this way the fitted
parameters enter only as multiplicative factors on specific component
histograms, $M_{ij}^{(k)}$, which allows us to use pre-made histograms 
to re-calculate 
expected yields,  
avoids the need to loop over the simulated events
at each step in the
$\chi^2$ minimization process and results in a dramatic
reduction in the required computation time.

\subsection{\boldmath Fit parameters and inputs}
\begin{table}[!htb]
\begin{center}
\caption{Input parameters for the fit.
\label{tab:inputParms}
}
\begin{tabular}{l|c} \hline\hline
 parameter & value \\\hline 
 $R_1$ & $1.429\pm0.061\pm0.044$ \\
 $R_2$ & $0.827\pm0.038\pm0.022$ \\
 $D^{**}$ FF slope & $-1.5\pm0.5$ \\
 $\mathcal{B}(B^{-} \to D^{(*)}\pi\ell\nub)$ & $0.0151 \pm 0.0015$ \\
 $f_{D_2^*/D_1}$ & $0.74 \pm 0.20$ \\
 $f_{D_0^* D\pi/D_1 D_2^*}$ & $0.87 \pm 0.43$ \\
 $f_{D_1' D^*\pi/D_1 D_2^*}$ & $0.36 \pm 0.30$ \\
 $f_{D\pi/D_0^*}$ & $0.21 \pm 0.21$ \\
 $f_{D^*\pi/D_1'}$ & $0.14 \pm 0.14$ \\
 $\mathcal{B}(B^-\to D^{(*)}\pi\pi\ell\nub)$ & $0.011 \pm 0.011$ \\
 $f_{D_2^*}$ & $1.7 \pm 0.4$ \\
 $\mathcal{B}(D^{*+} \to D^0 \pi^+)$ & $0.677 \pm 0.005$ \\
 $\mathcal{B}(D^0 \to K^- \pi^+)$ & $0.0389 \pm 0.0005$ \\
 $\mathcal{B}(D^+ \to K^- \pi^+\pi^+)$ & $0.0922 \pm 0.0021$ \\
 $\tau_{B^-} / \tau_{B^0}$ & $1.071 \pm 0.009$ \\
 $f_{+-} / f_{00}$ & $1.065 \pm 0.026$ \\
\hline\hline
\end{tabular}
\end{center}
\end{table}
The semileptonic decay widths of $\Bb\to D\ell\nub$, $\Bb\to\Dstar\ell\nub$
and $\Bb\to D^{**}\ell\nub$ are required to be equal for $B^+$ and $B^0$.  
We also require isospin invariance in the
decays $D^{**}\to D^{(*)}\pi$.
As a result, the $C_j$ depend on the following quantities: 
${\mathcal B}(\Bm\to \Dz\ell\nub)$ and form factor slope $\rho_D^2$ 
for $\Bb\to D\ell\nub$ and ${\mathcal B}(\Bm\to \Dstarz\ell\nub)$
and form-factor parameters $R_1$, $R_2$ 
and $\rho^2_{\Dstar}$ for $\Bb\to \Dstar\ell\nub$.
We fix $R_1$ and $R_2$ to the values obtained in
Ref.~\cite{ref:BaBarDstarlnu}.
The background contributions are kept at the values determined in
the simulation.
The overall normalizations of the
$\Bb\to D^{(*)}\pi\ell\nub$ and
$\Bb\to D^{(*)}\pi\pi\ell\nub$ components are also fixed. 
For the relevant $D$ decay branching fractions 
we use the values from Ref.~\cite{ref:PDG2008}.
The values of input parameters are listed in Table~\ref{tab:inputParms},
where 
$f_{D_2^*}$ is defined as the ratio $\mathcal{B}(D_2^{*+} \to
D^0 \pi^+) /\mathcal{B}(D_2^{*+} \to D^{*0} \pi^+)$~\cite{ref:PDG2008,ref:BtoNarrowDsslnu}, and
$f_{+-}/f_{00}$ is the ratio of branching fractions
$\mathcal{B}(\FourS\to\BpBm)/\mathcal{B}(\FourS\to\BzBzb)$~\cite{ref:PDG2008}.
All fixed values are varied in assessing systematic uncertainties.
\begin{table}
\caption{\label{XcStates} Assumed masses, widths, and spins of the 
four hypothetical high-mass states contributing to 
$\Bb\to D^{(*)}\pi\pi\ell\nub$ decays.}
\begin{tabular}{lccc} \hline\hline\noalign{\vskip 1pt}
    name   & mass ($\gev/c^2$)  & width ($\gev$)
& spin \\ \hline
     $X_c$    & 2.61  & 0.3 & 0 \\ 
     $X_c^*$  & 2.61  & 0.3 & 1 \\ 
     $Y_c$    & 2.87  & 0.1 & 0 \\ 
     $Y_c^*$  & 2.87  & 0.1 & 1 \\ 
\hline\hline
\end{tabular}
\end{table}

\subsection{\boldmath Fit results}
\label{subsec:fitresults}
\begin{table*}[htbp]
\begin{center}
\caption{Fit results on the electron and muon samples, and their
combination.  The first error is statistical, the second, systematic.
\label{tab:fitresults}
}
\begin{tabular}{lccc} \hline\hline
  Parameters & $De$ sample & $D\mu$ sample & combined result \\ \hline\noalign{\vskip 1pt}

  $\rho_D^2$ & $1.23 \pm 0.05 \pm 0.08$ & $1.13\pm 0.07\pm 0.09$ 
             & $\DSlope \pm \DSlopeStE \pm \DSlopeSyE$ \\ 
  $\rho_{D^*}^2$   & $1.23 \pm 0.02 \pm 0.07$ & $1.24\pm 0.03\pm 0.07$
             & $\DsSlope \pm \DsSlopeStE \pm \DsSlopeSyE$ \\ 
  $\mathcal{B}( D^{0} \ell \nub)$(\%) 
             & $2.38 \pm 0.03 \pm 0.14$ & $2.26\pm 0.04\pm 0.16$
             & $\DBF \pm \DBFStE \pm \DBFSyE$ \\ 
  $\mathcal{B}( D^{*0} \ell \nub)$(\%) 
             & $5.45 \pm 0.03 \pm 0.22$ & $5.27\pm 0.04\pm 0.37$ 
             & $\DsBF \pm \DsBFStE \pm \DsBFSyE$ \\ \hline\noalign{\vskip 1pt}
  $\chi^2$/n.d.f. (probability) & 422/470 (0.94) & 494/467 (0.19) & 2.2/4 (0.71) \\ 
\hline\hline
\end{tabular}
\end{center}
\end{table*}
\begin{table}
\begin{center}
\caption{Statistical correlation coefficients between parameters from the
fits to the electron and muon samples.
\label{tab:fitstatcorr}
}
\begin{tabular}{l|ccc|ccc} \hline\hline
 & \multicolumn{3}{c|}{$De$ sample}  & \multicolumn{3}{c}{$D\mu$ sample} \\  & $\rho_D^2$ & $\rho_{D^*}^2$ & $\mathcal{B}(D)$  
   & $\rho_D^2$ & $\rho_{D^*}^2$ & $\mathcal{B}(D)$ \\ \hline\noalign{\vskip 1pt}
$\rho_{D^*}^2$ &
$-0.299$ & & & $-0.302$ & &  \\
$\mathcal{B}(D)$ &
$+0.307$ & $+0.180$ & & $+0.279$ & $+0.198$ & \\
$\mathcal{B}(D^*)$ &
$-0.388$ & $+0.075$ & $-0.526$ & $-0.396$ & $+0.069$ & $-0.519$ \\
\hline\hline
\end{tabular}
\end{center}
\end{table}

The fit is performed separately on the electron and muon samples.
The results of these fits are given in Table~\ref{tab:fitresults}.
Both fits give good $\chi^2$ probabilities.  
The corresponding $\Bzb$ branching fractions are obtained from the
$\Bm$ results by dividing by the lifetime ratio~\cite{ref:PDG2008} 
$\tau_{\Bm}/\tau_{\Bzb}=1.071$.  
The statistical correlations for the electron and muon samples
are given in Table~\ref{tab:fitstatcorr}.
Fig.~\ref{fig:1d}  and Fig.~\ref{fig:1dMu}
show the projected distributions on the three kinematic variables for
the electron and muon samples along with the ratio of data over fit.

The results of the separate fits to the $De$ and $D\mu$ 
samples are combined using the full $8\times 8$ covariance matrix.
This matrix is built from a block-diagonal statistical covariance matrix,
with one $4\times 4$ block coming from the fit to each lepton sample,
and the full $8\times 8$ systematic covariance matrix described in
 Sec.~\ref{sec:systematics}.  The systematic covariance matrix
consists of $4\times 4$ matrices for the electron and muon
parameters and a $4\times 4$ set of electron-muon covariance terms.
The corresponding correlation coefficients are given in
Table~\ref{tab:emusyscorr}.
There is an advantage to combining the electron and muon results 
after the systematic errors have been evaluated; the results are 
weighted optimally ({\it e.g.}, the difference in lepton identification
efficiency uncertainties is taken into account) and the $\chi^2$
from the combination provides a valid measure of the compatibility
of the electron and muon results.  The combined results are given
in Table~\ref{tab:fitresults}, and the 
correlation coefficients corresponding to 
the combined statistical and systematic errors are given 
in Table~\ref{tab:combinedcorr}.

\begin{table}
\begin{center}
\caption{Correlation coefficients for systematic errors. 
The upper and lower
diagonal blocks correspond to electrons and muons, respectively.\label{tab:emusyscorr}
}
\begin{tabular}{l|cccc|ccc} \hline\hline
 & \multicolumn{4}{c|}{$De$ sample}  & \multicolumn{3}{c}{$D\mu$ sample} \\  & $\rho_D^2$ & $\rho_{D^*}^2$ & $\mathcal{B}(D)$ & $\mathcal{B}(D^{*})$  
   & $\rho_D^2$ & $\rho_{D^*}^2$ & $\mathcal{B}(D)$ \\ \hline\noalign{\vskip 1pt}
$\rho_{D^*}^2$ &
$-0.02$ & & & & & & \\
$\mathcal{B}(D)$ &
$+0.74$ & $+0.08$ & & & & & \\
$\mathcal{B}(D^*)$ &
$-0.22$ & $+0.36$ & $+0.31$ & & & & \\ \hline\noalign{\vskip 1pt}
$\rho_D^2$ &
$+0.75$ & $-0.17$ & $+0.47$ & $-0.35$ & & & \\
$\rho_{D^*}^2$ &
$-0.07$ & $+0.98$ & $+0.02$ & $+0.31$ & $-0.15$ & & \\
$\mathcal{B}(D)$ &
$+0.46$ & $+0.00$ & $+0.64$ & $+0.17$ & $+0.16$ & $-0.03$ & \\
$\mathcal{B}(D^*)$ &
$-0.17$ & $+0.19$ & $+0.12$ & $+0.54$ & $-0.48$ & $+0.17$ & $+0.67$ \\
\hline\hline
\end{tabular}
\end{center}
\end{table}

\begin{table}[!htb]
\begin{center}
\caption{Output correlation matrix for combined samples.\label{tab:combinedcorr}
}
\begin{tabular}{l|ccc} \hline\hline\noalign{\vskip 1pt}
& $\rho_D^2$ & $\rho_{D^*}^2$ & $\mathcal{B}(D)$ \\\hline \noalign{\vskip 1pt}
$\rho_{D^*}^2$ &
$-0.129$ & & \\
$\mathcal{B}(D)$ &
$+0.609$ & $+0.023$ & \\
$\mathcal{B}(D^*)$ &
$-0.285$ & $+0.308$ & $+0.283$ \\
\hline\hline
\end{tabular}
\end{center}
\end{table}

\subsection{\boldmath Fit validation}
\label{subsec:validation}
The fit was validated in several ways.  A large number of simulated
experiments were generated based on random samples drawn from the
histograms used in the fit.  The fit was performed on these simulated
experiments to check for biases in the fitted values or associated
variances.  Small biases in the fitted values of several parameters -
in no case exceeding 0.1 standard deviations 
for both electron and muon samples
- were found.  Given the
smallness of the biases we do not correct the fit results.
Additional sets of simulated experiments were generated with
alternative values for the parameters.  In each case the fit
reproduced the alternative values within statistical uncertainties.  An
independent 
sample of fully-simulated events was also used to validate the fit.

Additional fits were performed on the data 
to look for inconsistencies and
quantify the impact of additional constraints.  The electron and
muon samples were combined before fitting; the results 
were compatible with expectations.
Data samples collected in different years were fitted separately;
the fit results agree within statistical uncertainties.
The minimum number of expected entries per bin was varied
from 10 to 100; the impact on the fitted parameters was negligible.
Different binnings in the variables $p_\ell^*$, $p_D^*$ and 
$\cos\theta_{B-D\ell}$ were tried; the fit results were in each case
consistent with the nominal values.  The boundaries of the
$D$ mass peak and sideband regions were varied by $\pm 2$~MeV$/c^2$; the 
impact on the fitted parameters was negligible.

Additional fits were performed in which $R_1$ and $R_2$ 
were treated as free parameters. 
The results, including associated systematic uncertainties, 
are given in Table~\ref{tab:fitresultsR1R2}.
\begin{table*}[htbp]
\begin{center}
\caption{Results on the electron, muon and combined samples 
when fitting $R_1$ and $R_2$.
\label{tab:fitresultsR1R2}
}
\begin{tabular}{lccc} \hline\hline
  Parameters & $De$ sample & $D\mu$ sample & combined result \\ \hline\noalign{\vskip 1pt}
  $\rho_D^2$     & $1.22 \pm 0.05 \pm 0.10$ & $1.10 \pm 0.07 \pm 0.10 $ 
                 & $1.16 \pm 0.04 \pm 0.08$ \\ 
  $\rho_{D^*}^2$ & $1.34 \pm 0.05 \pm 0.09$ & $1.33 \pm 0.06 \pm 0.09 $ 
                 & $1.33 \pm 0.04 \pm 0.09$ \\ 
  $R_1$          & $1.59 \pm 0.09 \pm 0.15$ & $1.53 \pm 0.10 \pm 0.17 $
                 & $1.56 \pm 0.07 \pm 0.15$ \\ 
  $R_2$          & $0.67 \pm 0.07 \pm 0.10$ & $0.68 \pm 0.08 \pm 0.10 $
                 & $0.66 \pm 0.05 \pm 0.09$ \\ 
  $\mathcal{B}( D^{0} \ell \nub)$(\%) 
                 & $2.38 \pm 0.04 \pm 0.15$ & $2.25 \pm 0.04 \pm 0.17 $
                 & $2.32 \pm 0.03 \pm 0.13$ \\ 
  $\mathcal{B}( D^{*0} \ell \nub)$(\%) 
                 & $5.50 \pm 0.05 \pm 0.23$ & $5.34 \pm 0.06 \pm 0.37 $ 
                 & $5.48 \pm 0.04 \pm 0.22$ \\ \hline\noalign{\vskip 1pt}
  $\chi^2$/n.d.f. (probability) & 416/468 (0.96) & 488/464 (0.21) & 2.0/6 (0.92) \\ 
\hline\hline
\end{tabular}
\end{center}
\end{table*}
\begin{table}[htb]
\begin{center}
\caption{Output correlation coefficients for combined samples with
$R_1$ and $R_2$ fitted.
\label{tab:combinedcorrR1R2}
}
\begin{tabular}{l|ccccc} \hline\hline\noalign{\vskip 1pt}
& $\rho_D^2$ & $\rho_{D^*}^2$ & $R_1$ & $R_2$ 
& $\mathcal{B}(D)$ \\
\hline \noalign{\vskip 1pt}
$\rho_{D^*}^2$ &
$-0.435$  & & & & \\
$R_1$ &
$-0.252$ & $+0.752$  & & & \\
$R_2$ &
$+0.519$ & $-0.787$ & $-0.740$  & & \\
$\mathcal{B}(D)$ &
$+0.602$ & $-0.056$ & $+0.114$ & $+0.102$  & \\
$\mathcal{B}(D^*)$ &
$-0.310$ & $+0.406$ & $+0.139$ & $-0.309$ & $+0.212$  \\
\hline\hline
\end{tabular}
\end{center}
\end{table}
Correlation coefficients for the combined fit are given in 
Table~\ref{tab:combinedcorrR1R2}.
The three $D^*$ form factor parameters are highly correlated.
Comparing this set of parameters with 
the previous measurement~\cite{ref:BaBarDstarlnu}, we find
they are consistent at the 36\%\ C.L.

\section{\boldmath Systematic studies}
\label{sec:systematics}
There are several sources of systematic uncertainty in this analysis.
Table~\ref{SysErrors01} summarizes the systematic 
uncertainties on the quantities of interest; 
these were used in determining the systematic errors
and correlations given in Tables~\ref{tab:fitresults} 
and \ref{tab:emusyscorr}.

The parameters $R_1$ and $R_2$ 
are varied taking their correlation ($-0.84$) into account.
We transform $R_1$ and $R_2$ into a set of parameters 
$R_1'$ and $R_2'$ that diagonalize the error matrix, and vary
$R_1'$ and $R_2'$ independently.
The $D^{**}$ form factor shape is varied in two ways: the slope is varied
from $-2.0$ to $-1.0$, and the approximation B$_1$ from
Ref.~\cite{ref:LLSW} is replaced with B$_2$ 
(see also Appendix~\ref{sec:LLSWmodel}).  The total and relative
branching fractions of the $D^{**}$ components in $\Bb\to
D^{(*)}\pi\ell\nub$ decays are varied independently using the values
in Table~\ref{tab:inputParms}.
The $D^*/D$ ratio of non-resonant decays, 
which is defined by $\mathcal{B}^{NR}(\Bb \to  D^* \pi\ell \nub)
/\mathcal{B}^{NR}(\Bb \to  D \pi\ell \nub)$,
is 0.3 in the nominal fit; we vary the ratio from 0.1 to 1.0.  
The branching fraction of $\Bb\to D^{(*)}\pi\pi\ell\nub$
decays is varied as given
in Table~\ref{tab:inputParms}, and the production
ratios for the states used to model $\Bb\to D^{(*)}\pi\pi\ell\nub$
decays, $X_c^*/X_c$, $Y^*_c/Y_c$, $X_c/Y_c$ and $X^*_c/Y^*_c$, 
are varied independently from 0.5 to 2.0.
To evaluate the effect of $D_1 \to D\pi\pi$
decays~\cite{ref:BELLE_D1Dpipi}, one half of the
$\Bb\to D^{(*)}\pi\pi\ell\nub$ component is replaced by $D_1 \to D\pi\pi$
decays; the differences in fitted values are taken as systematic
uncertainties.
  
The other parameters listed in Table~\ref{tab:inputParms} are also varied 
within their uncertainties.  
The determination of the number of $\BB$ events introduces a 
normalization uncertainty of 1.1\% on the branching fractions.  
The uncertainty in the luminosity ratio between on-peak and off-peak 
data is 0.25\%.

The $B$ momentum distribution is determined from the well-measured beam
energy and $B^0$ mass. 
The uncertainty of 0.2 MeV in the beam energy measurement leads to
a systematic error. 
Uncertainties arising from the simulation of 
the detector response to charged particle
reconstruction and particle identification are studied by varying the
efficiencies and mis-identification probabilities based on 
comparisons between data and simulation on dedicated control samples.  
The uncertainty arising
from radiative corrections is studied by comparing the results using
PHOTOS~\cite{ref:photos} to simulate final state radiation 
(default case) with those obtained with
PHOTOS turned off. We take 25 \%\ of the difference as an error.  
The uncertainty in the simulation of bremsstrahlung is based on an
understanding of the detector material from studies of photon conversions
and hadronic interactions.
The uncertainty associated with the charge particle vertex
requirements for the $D$ and $B$ decay points
is evaluated by loosening the vertex probability cuts.  
The uncertainty arising from $\BzBzb$ mixing is negligible.

Branching fractions
in background simulations
are varied within their measured
uncertainties~\cite{ref:PDG2008}.  
The inclusive differential branching fractions versus $D$
momentum for $B$ meson decays to $\Dz$, $\Dzb$, $\Dp$ and $\Dm$ mesons,
which affect some background components,
are varied using the measurements from
Ref.~\cite{ref:BaBarInclusiveBtoD}.  

The overall covariance matrix for the 8 fitted quantities (4 electron and
4 muon parameters) is built from the individual systematic variations
as follows.  For each variation 
taken, an 8-component vector $\Delta\vec{\alpha}$
of parameter differences 
between the alternative fit and the nominal fit is recorded.  The
$ij$ element of the systematic error covariance matrix is
the sum over all systematic variations $k$:
\begin{equation}
  {V_{\mathrm{sys}}}_{ij} = \sum_k 
  \Delta\alpha_i^{(k)} \Delta\alpha_j^{(k)}.
\end{equation}
The corresponding correlation matrix is given in Table~\ref{tab:emusyscorr}.

\begin{table*}
\begin{center}
\caption{\label{SysErrors01} Systematic uncertainties on fitted parameters,
given in \%.  Numbers are negative when the fitted value decreases as input 
parameter increases.}
\begin{tabular}{l|rrrrrr|rrrrrr} \hline\hline
 & \multicolumn{6}{c|}{Electron sample}  & \multicolumn{6}{c}{Muon sample} \\
 item 
& $\rho_D^2$ & $\rho_{D^*}^2$ 
& $\mathcal{B}(D\ell\nub)$& $\mathcal{B}(D^{*}\ell\nub)$
& $\mathcal{G}(1)\Vcb$ & $\mathcal{F}(1)\Vcb$
& $\rho_D^2$ & $\rho_{D^*}^2$
& $\mathcal{B}(D\ell\nub)$& $\mathcal{B}(D^{*}\ell\nub)$
& $\mathcal{G}(1)\Vcb$ & $\mathcal{F}(1)\Vcb$
 \\ \hline\noalign{\vskip 1pt}
 $R_1'$ & $0.44$ & $2.74$ & $0.71$ & $-0.38$ & $0.60$ & $0.71$ & $0.50$ & $2.67$ & $0.74$ & $-0.40$ & $0.63$ & $0.70$ \\ 
 $R_2'$ & $-0.40$ & $1.02$ & $-0.18$ & $0.30$ & $-0.32$ & $0.49$ & $-0.45$ & $0.96$ & $-0.19$ & $0.30$ & $-0.33$ & $0.48$ \\ 
 $D^{**}$ slope & $-1.42$ & $-2.52$ & $-0.07$ & $-0.09$ & $-0.82$ & $-0.87$ & $-1.42$ & $-2.58$ & $-0.10$ & $-0.10$ & $-0.77$ & $-0.92$ \\ 
 $D^{**}$ FF approximation & $-0.87$ & $0.33$ & $-0.12$ & $0.19$ & $-0.54$ & $0.20$ & $-0.99$ & $0.59$ & $-0.12$ & $0.21$ & $-0.59$ & $0.30$ \\ 
 $\mathcal{B}(B^- \rightarrow D^{(*)}\pi\ell\nub)$ & $0.28$ & $-0.27$ & $-0.22$ & $-0.80$ & $0.04$ & $-0.49$ & $0.59$ & $-0.32$ & $-0.13$ & $-0.86$ & $0.24$ & $-0.54$ \\ 
 $f_{D_2^*/D_1}$ & $-0.39$ & $0.16$ & $-0.38$ & $0.16$ & $-0.41$ & $0.13$ & $-0.50$ & $0.17$ & $-0.41$ & $0.18$ & $-0.47$ & $0.15$ \\ 
 $f_{D_0^*D\pi/D_1D_2^*}$ & $-2.30$ & $1.12$ & $-1.53$ & $0.97$ & $-2.07$ & $0.85$ & $-3.13$ & $1.23$ & $-1.53$ & $1.02$ & $-2.41$ & $0.93$ \\ 
 $f_{D_1'D^*\pi/D_1D_2^*}$ & $1.82$ & $-1.14$ & $1.30$ & $-0.65$ & $1.65$ & $-0.70$ & $2.44$ & $-1.15$ & $1.35$ & $-0.72$ & $1.91$ & $-0.75$ \\ 
 $f_{D\pi/D_0^*}$ & $-0.88$ & $-1.28$ & $0.36$ & $0.17$ & $-0.31$ & $-0.34$ & $-0.83$ & $-1.23$ & $0.31$ & $0.18$ & $-0.27$ & $-0.33$ \\ 
 $f_{D^*\pi/D_1'}$ & $-0.21$ & $-0.05$ & $-0.13$ & $0.21$ & $-0.18$ & $0.09$ & $-0.30$ & $-0.04$ & $-0.15$ & $0.23$ & $-0.23$ & $0.10$ \\ 
 NR $D^*/D$ ratio & $0.58$ & $-0.16$ & $0.11$ & $-0.09$ & $0.38$ & $-0.04$ & $0.66$ & $-0.16$ & $0.11$ & $-0.09$ & $0.40$ & $-0.03$ \\ 
 $\mathcal{B}(B^- \rightarrow D^{(*)}\pi\pi\ell\nub)$ & $1.19$ & $-1.97$ & $0.25$ & $-1.28$ & $0.78$ & $-1.28$ & $1.98$ & $-1.71$ & $0.40$ & $-1.20$ & $1.20$ & $-1.18$ \\ 
 $X^*/X$ and $Y^*/Y$ ratio & $0.61$ & $-1.15$ & $0.09$ & $-0.27$ & $0.39$ & $-0.52$ & $0.74$ & $-1.02$ & $0.08$ & $-0.24$ & $0.42$ & $-0.47$ \\ 
 $X/Y$ and $X^*/Y^*$ ratio & $0.76$ & $-0.83$ & $0.21$ & $-0.65$ & $0.52$ & $-0.60$ & $1.09$ & $-0.76$ & $0.25$ & $-0.63$ & $0.68$ & $-0.57$ \\ 
 $D_1\rightarrow D\pi\pi$ & $2.22$ & $-1.54$ & $0.74$ & $-1.08$ & $1.63$ & $-1.05$ & $2.74$ & $-1.48$ & $0.76$ & $-1.06$ & $1.81$ & $-1.03$ \\ 
 $f_{D_2^*}$ & $-0.14$ & $-0.01$ & $-0.10$ & $0.07$ & $-0.12$ & $0.03$ & $-0.16$ & $-0.01$ & $-0.10$ & $0.07$ & $-0.13$ & $0.03$ \\ 
 $\mathcal{B}(D^{*+}\rightarrow D^0\pi^+)$ & $0.73$ & $-0.01$ & $0.43$ & $-0.34$ & $0.62$ & $-0.17$ & $0.80$ & $-0.00$ & $0.41$ & $-0.33$ & $0.61$ & $-0.17$ \\ 
 $\mathcal{B}(D^{0}\rightarrow K^-\pi^+)$ & $0.69$ & $0.02$ & $-0.21$ & $-1.63$ & $0.29$ & $-0.80$ & $0.92$ & $0.12$ & $-0.27$ & $-1.68$ & $0.35$ & $-0.80$ \\ 
 $\mathcal{B}(D^{+}\rightarrow K^-\pi^+\pi^+)$ & $-1.46$ & $-0.42$ & $-2.17$ & $0.30$ & $-1.89$ & $0.01$ & $-1.43$ & $-0.42$ & $-2.10$ & $0.28$ & $-1.77$ & $-0.01$ \\ 
 $\tau_{B^-}/\tau_{B^0}$ & $0.26$ & $0.16$ & $0.63$ & $0.27$ & $0.46$ & $0.19$ & $0.22$ & $0.16$ & $0.58$ & $0.28$ & $0.41$ & $0.19$ \\ 
 $f_{+-}/f_{00}$ & $0.88$ & $0.43$ & $0.66$ & $-0.53$ & $0.82$ & $-0.12$ & $0.91$ & $0.48$ & $0.57$ & $-0.52$ & $0.75$ & $-0.10$ \\ 
 Number of $B\Bb$ events & $0.00$ & $-0.00$ & $-1.11$ & $-1.11$ & $-0.55$ & $-0.55$ & $0.00$ & $-0.00$ & $-1.11$ & $-1.11$ & $-0.55$ & $-0.55$ \\ 
 Off-peak Luminosity & $0.05$ & $0.01$ & $-0.02$ & $-0.00$ & $0.02$ & $0.00$ & $0.07$ & $0.00$ & $-0.02$ & $-0.00$ & $0.02$ & $-0.00$ \\ 
 $B$ momentum distrib. & $-0.96$ & $0.63$ & $1.29$ & $-0.54$ & $-1.15$ & $0.48$ & $1.30$ & $-0.10$ & $1.27$ & $-0.64$ & $1.31$ & $-0.35$ \\ 
 Lepton PID eff & $0.52$ & $0.16$ & $1.21$ & $0.82$ & $0.90$ & $0.46$ & $3.30$ & $0.06$ & $5.11$ & $5.83$ & $1.99$ & $2.90$ \\ 
 Lepton mis-ID & $0.03$ & $0.01$ & $-0.01$ & $-0.01$ & $0.01$ & $-0.00$ & $2.65$ & $0.70$ & $-0.59$ & $-0.50$ & $1.06$ & $-0.01$ \\ 
 Kaon PID & $0.07$ & $0.80$ & $0.28$ & $0.23$ & $0.18$ & $0.38$ & $1.02$ & $0.71$ & $0.35$ & $0.29$ & $0.70$ & $0.39$ \\ 
 Tracking eff & $-1.02$ & $-0.43$ & $-3.35$ & $-2.00$ & $-2.25$ & $-1.15$ & $-0.63$ & $-0.28$ & $-3.37$ & $-2.09$ & $-2.02$ & $-1.14$ \\ 
 Radiative corrections & $-3.13$ & $-1.04$ & $-2.87$ & $-0.74$ & $-3.02$ & $-0.71$ & $-0.76$ & $-0.61$ & $-0.82$ & $-0.25$ & $-0.79$ & $-0.33$ \\ 
 Bremsstrahlung & $0.07$ & $0.00$ & $-0.13$ & $-0.28$ & $-0.04$ & $-0.14$ & $0.00$ & $0.00$ & $0.00$ & $0.00$ & $0.00$ & $0.00$ \\ 
 Vertexing & $0.83$ & $-0.64$ & $0.63$ & $0.60$ & $0.78$ & $0.09$ & $1.79$ & $-0.76$ & $0.97$ & $0.54$ & $1.41$ & $0.01$ \\ 
Background total & $1.39$ & $1.12$ & $0.64$ & $0.34$ & $1.07$ & $0.51$ & $1.58$ & $1.09$ & $0.67$ & $0.38$ & $1.16$ & $0.49$ \\ \hline
 {\bf Total} & {\bf 6.25} & {\bf 5.66} & {\bf 6.01} & {\bf 4.03} & {\bf 5.99} & {\bf 3.20} & {\bf 8.12} & {\bf 5.47} & {\bf 7.35} & {\bf 7.07} & {\bf 6.06} & {\bf 4.23} \\ 
\hline\hline
\end{tabular}
\end{center}
\end{table*}

\section{\boldmath Determination of \Vcb}
\label{sec:physics}
The combined fit results with their full covariance matrix
are used to calculate ${\mathcal G}(1)\Vcb$ and ${\mathcal F}(1)\Vcb$:
\begin{eqnarray}
{\mathcal G}(1)\Vcb &=& (\GVcb \pm \GVcbStE \pm \GVcbSyE) \times 10^{-3} \\
{\mathcal F}(1)\Vcb &=& (\FVcb \pm \FVcbStE \pm \FVcbSyE) \times 10^{-3}.
\end{eqnarray}
The errors are statistical and systematic, respectively.
The associated correlations are $\corGSlope$ (between ${\mathcal G}(1)\Vcb$ and $\rho_D^2)$,
$\corFSlope$ (${\mathcal F}(1)\Vcb$ and $\rho_{D^*}^2$) and
$\corGF$ (${\mathcal G}(1)\Vcb$ and ${\mathcal F}(1)\Vcb$).

Using the values of $\mathcal{F}(1)\Vcb$ and $\mathcal{G}(1)\Vcb$
given above along with calculations of the form factor
normalizations allows one to determine $\Vcb$.  Using a recent
lattice QCD calculation, 
$\mathcal{G}(1)=1.074\pm 0.018\pm 0.016$~\cite{FermiLabMilk},
multiplied by the electroweak correction~\cite{QEDCorr} of 1.007, 
we find
\begin{eqnarray}
D\ell\nu:\ \Vcb &=& (\VcbG \pm \VcbGStE \pm \VcbGSyE \pm \VcbGThE) \times 10^{-3}.
\end{eqnarray}
where the errors are statistical, systematic and theoretical, respectively.
For $\Bb\to D^*\ell\nub$ we use a lattice QCD calculation of the form factor,
$\mathcal{F}(1)=0.921 \pm 0.013 \pm 0.020$~\cite{Hashimoto:2001nb}, 
along with the electroweak correction factor, 
to find
\begin{eqnarray}
D^{*}\ell\nu:\ \Vcb &=& (\VcbF \pm \VcbFStE \pm \VcbFSyE \pm \VcbFThE) \times 10^{-3}.
\end{eqnarray}

The fits with $R_1$ and $R_2$ as free parameters give
\begin{eqnarray}
 {\mathcal G}(1)\Vcb &=& (42.8 \pm 0.9 \pm 2.3) \times 10^{-3} \\
 {\mathcal F}(1)\Vcb &=& (35.6 \pm 0.3 \pm 1.0) \times 10^{-3},
\end{eqnarray}
with correlation coefficients 
$+0.92$ (between ${\mathcal G}(1)\Vcb$ and $\rho_D^2)$, 
$+0.41$ (${\mathcal F}(1)\Vcb$ and $\rho_{D^*}^2$) and 
$-0.03$ (${\mathcal G}(1)\Vcb$ and ${\mathcal F}(1)\Vcb$).

\section{Discussion}
\label{sec:discussion}

The branching fractions and slope parameters measured here for $\Bb\to
D\ell\nub$ and $\Bb\to \Dstar\ell\nub$ are consistent with the world
averages~\cite{ref:HFAG} for these quantities.  The measurements of
$\rho^2_D$ and $\mathcal{G}(1)\Vcb$ represent significant improvements on
existing knowledge.  
The experimental technique used here, namely a simultaneous global fit to
$\Bb\to \Dz X\ell\nub$ and $\Bb\to \Dp X\ell\nub$ combinations,
is complementary to previous measurements.  In particular, it
does not rely on the reconstruction of the soft transition pion
from the $\Dstar\to D\pi$ decay.  

The results obtained here,
which are given in Table~\ref{tab:fitresults}, 
can be combined with the existing \babar\ measurements  
listed in Table~\ref{tab:previousresults}.
For
$\Bb\to\Dstar\ell\nub$, we combine the present results with two \babar\
measurements of $\rho_{\Dstar}^2$ and
$\mathcal{F}(1)\Vcb$~\cite{ref:BaBarDstarlnu,ref:BaBarDstarzlnu} and
four measurements of
$\mathcal{B}(\Bb\to\Dstar\ell\nub)$\cite{ref:BaBarDstarlnu,ref:BaBarDstarzlnu,ref:BaBarbclnutagged}.
We neglect the tiny statistical correlations among the measurements and
treat the systematic uncertainties as fully correlated within a given
category (background, detector modeling, etc.).
We assume the semileptonic decay widths of $B^+$ and $B^0$ 
to be equal 
and adjust all measurements to the values
of the $\FourS$ and $D$ decay branching fractions used in this article
to obtain
\begin{eqnarray}
{\mathcal B}(\Bm\to\Dstarz\ell\nub) &=& (5.49\pm 0.19)\% \\
{\rho^2_{\Dstar}} &=& 1.20\pm 0.04 \\
{\mathcal F}(1)\Vcb &=& (34.8 \pm 0.8) \times 10^{-3}.
\end{eqnarray}
The associated $\chi^2$ probabilities of the averages are $0.39$, 
$0.86$ and $0.27$, respectively.  The average of the
$\mathcal{B}(\Bb\to D\ell\nub)$ result with the two existing
\babar\ measurements~\cite{ref:BaBarbclnutagged} is
\begin{equation}
{\mathcal B}(\Bm\to\Dz\ell\nub) = (2.32\pm 0.09)\%
\end{equation}
with a $\chi^2$ probability of $0.88$.
\begin{table*}[htbp]
\begin{center}
\caption{Previously published \babar\ 
results~\cite{ref:BaBarDstarlnu,ref:BaBarDstarzlnu,ref:BaBarbclnutagged}.
\label{tab:previousresults}
}
\begin{tabular}{l|ccc} \hline\hline\noalign{\vskip 1pt}
  Parameters & Ref.~\cite{ref:BaBarDstarlnu} & Ref.~\cite{ref:BaBarDstarzlnu} 
             & Ref.~\cite{ref:BaBarbclnutagged} \\ \hline\noalign{\vskip 1pt}
  $\rho_{D^*}^2$   
             & $1.191 \pm 0.048 \pm 0.028$ & $1.16\pm 0.06\pm 0.08$
             &  \\ 
  $\mathcal{B}(\Bm\to D^{*0} \ell \nub)$(\%) 
             &                          & $5.56\pm 0.08\pm 0.41$ 
             & $5.83 \pm 0.15 \pm 0.30$ \\ 
  $\mathcal{B}(\Bzb\to D^{*+} \ell \nub)$(\%) 
             & $4.69 \pm 0.04 \pm 0.34$ &  
             & $5.49 \pm 0.16 \pm 0.25$ \\
  ${\mathcal F}(1)\Vcb$ ($\times 10^{-3}$) 
             & $34.4 \pm 0.3 \pm 1.1$   & $35.9\pm 0.6\pm 1.4$
             &  \\ \hline
  $\mathcal{B}(\Bm\to D^{0} \ell \nub)$(\%) 
             &  & 
             & $2.33 \pm 0.09 \pm 0.09$ \\ 
  $\mathcal{B}(\Bzb\to D^{+} \ell \nub)$(\%) 
             &  & 
             & $2.21 \pm 0.11 \pm 0.12$ \\ 
\hline\hline
\end{tabular}
\end{center}
\end{table*}

The simultaneous measurements of $\mathcal{G}(1)\Vcb$ and 
$\mathcal{F}(1)\Vcb$ allow a determination
of the ratio $\mathcal{G}(1)/\mathcal{F}(1)$ which can be compared
directly with theory.  We find
\begin{eqnarray}
\mathrm{Measured}: &{\mathcal{G}(1)}/{\mathcal{F}(1)}& 
                                             = \GFRatio \pm \GFRatioE \\
\mathrm{Theory}:   &{\mathcal{G}(1)}/{\mathcal{F}(1)}& = 1.17\pm 0.04,
\end{eqnarray}
where we have assumed the theory errors on 
$\mathcal{F}(1)$~\cite{Hashimoto:2001nb}
and $\mathcal{G}(1)$~\cite{FermiLabMilk} to be independent.  
The measured ratio is consistent with the predicted ratio.

The excellent description obtained in this fit, at the $1\%$ statistical
level, of the dominant Cabibbo-favored semileptonic decays will
facilitate the determination of decay rates of Cabibbo-suppressed decays
over a larger kinematic region than has been feasible to date. This will
result in a reduction in the theoretical uncertainty on the
determination of $\Vub$.

To summarize: we use a global fit to $\Dz\ell$ and $\Dp\ell$ combinations
to measure the form factor parameters
\begin{eqnarray}
  \rho_D^2 &=& \DSlope \pm \DSlopeStE \pm \DSlopeSyE \nonumber \\
  \rho_{D^*}^2 &=& \DsSlope \pm \DsSlopeStE \pm \DsSlopeSyE,
\end{eqnarray}
in the commonly used HQET-based parameterization~\cite{ref:CLN}
and the branching fractions
\begin{eqnarray}
  \mathcal{B}(B^- \to  D^{0} \ell \nub) 
  &=& (\DBF \pm \DBFStE \pm \DBFSyE) \%\hspace{0.3cm} \nonumber \\
  \mathcal{B}(B^- \to  D^{*0} \ell \nub)
  &=& (\DsBF \pm \DsBFStE \pm \DsBFSyE) \%,
\end{eqnarray}
where the first error is statistical and the second systematic. 
The fit assumes the semileptonic decay widths of $B^+$ and $B^0$ 
to be equal.
These results are consistent with previous \babar\  
measurements~\cite{ref:BaBarDstarlnu,ref:BaBarDstarzlnu,ref:BaBarbclnutagged}.
From these slopes and branching fractions we determine
\begin{eqnarray}
  \mathcal{G}(1)|V_{cb}| 
  &=& (\GVcb \pm \GVcbStE \pm \GVcbSyE)\times10^{-3}\nonumber \\
  \mathcal{F}(1)|V_{cb}| 
  &=& (\FVcb \pm \FVcbStE \pm \FVcbSyE)\times10^{-3} .
\end{eqnarray}
The $\mathcal{G}(1)|V_{cb}|$ value is twice as precise as the current
world average.  The precision on $\mathcal{F}(1)|V_{cb}|$ 
equals that of the best single measurement, while coming from a 
complementary technique.  
From these results, we extract two values for $|V_{cb}|$:
\begin{eqnarray}
D^{*}\ell\nu:\ \Vcb &=& (\VcbF \pm \VcbFStE \pm \VcbFSyE \pm \VcbFThE) \times 10^{-3} \nonumber \\
D\ell\nu:\ \Vcb &=& (\VcbG \pm \VcbGStE \pm \VcbGSyE \pm \VcbGThE) \times 10^{-3},
\end{eqnarray}
where the errors correspond to statistical, systematic and theoretical
uncertainties, respectively.

\section{Acknowledgements}
\label{sec:Acknowledgments}
We are grateful for the 
extraordinary contributions of our \pep2\ colleagues in
achieving the excellent luminosity and machine conditions
that have made this work possible.
The success of this project also relies critically on the 
expertise and dedication of the computing organizations that 
support \babar.
The collaborating institutions wish to thank 
SLAC for its support and the kind hospitality extended to them. 
This work is supported by the
US Department of Energy
and National Science Foundation, the
Natural Sciences and Engineering Research Council (Canada),
the Commissariat \`a l'Energie Atomique and
Institut National de Physique Nucl\'eaire et de Physique des Particules
(France), the
Bundesministerium f\"ur Bildung und Forschung and
Deutsche Forschungsgemeinschaft
(Germany), the
Istituto Nazionale di Fisica Nucleare (Italy),
the Foundation for Fundamental Research on Matter (The Netherlands),
the Research Council of Norway, the
Ministry of Education and Science of the Russian Federation, 
Ministerio de Educaci\'on y Ciencia (Spain), and the
Science and Technology Facilities Council (United Kingdom).
Individuals have received support from 
the Marie-Curie IEF program (European Union) and
the A. P. Sloan Foundation.

\appendix

\section{Modeling of $\Bb\to D^{**}\ell\nul$ decays}
\label{sec:LLSWmodel}

The differential decay rates of $\Bb\to D^{**}\ell\nul$ decays
are given as functions of
$w$ and $\theta$~\cite{ref:LLSW}.  
This $\theta$ is the angle between the charged
lepton and the charmed meson in the rest frame of the virtual $W$
boson. Thus $\theta$ is related to $\theta_{\ell}$, which is defined in
Fig.~\ref{DecayGeom}, such that
\begin{equation}
  \cos\theta = \cos(\pi - \theta_{\ell}) = - \cos\theta_{\ell}.
\end{equation}
In the following subsections, we use the same notation as above, $r$ and $R$,
for the mass ratios of all four $D^{**}$ mesons. However, it is
implied that these are the ratios taken with corresponding charmed
meson masses. The following notations are also used in
the form factor formulae in the following subsections :
\begin{equation}
  \varepsilon_b \equiv \frac{1}{2m_b},
  \hspace{0.3cm}
  \varepsilon_c \equiv \frac{1}{2m_c}
\end{equation}
and
\begin{equation}
\begin{array}{l}
  \bar{\Lambda}= \textrm{ energy of the ground state doublet ($D$ and $D^*$)} \\
  \bar{\Lambda}'= \textrm{ energy of the excited $\frac{3}{2}^+$ doublet ($D_1$ and $D_2^*$)} \\
  \bar{\Lambda}^*= \textrm{ energy of the excited $\frac{1}{2}^+$ doublet ($D_0^*$ and $D_1'$)}.
\end{array}
\end{equation}

\subsubsection{$\Bb\to D_1 \ell \nub$}
\noindent The differential decay rate is given by
\begin{equation}
  \frac{d^2 \Gamma_{D_1}}{d w\, d\cos\theta}
   =  \Gamma_z 
    r^3 (w^2 - 1)^{1/2} \hspace{0.1cm}
   \mathcal{I}_{D_1} (w, \theta),
\end{equation}
where $\Gamma_z \equiv \frac{G_F^2 |V_{cb}|^2 m_B^5}{64\pi^3}$ and
\begin{eqnarray}
   \mathcal{I}_{D_1}(w, \theta) \hspace{6.2cm}\nonumber\\
   = (1-\cos^2\theta) [(w - r) f_{V_1} 
     + (w^2 - 1)(f_{V_3} + r f_{V_2})]^2   \nonumber\\
   + (1 - 2 r w + r^2)
       \lbrack(1 + \cos^2\theta)(f_{V_1}^2 + (w^2 - 1)f_{A}^2) \nonumber\\
      -4\cos\theta \sqrt{w^2 - 1} f_{V_1} f_{A}\rbrack \hspace{0.1cm}
\end{eqnarray}
and $f_{V_1}(w)$, $f_{V_2}(w)$, $f_{V_3}(w)$ and $f_{A}(w)$ are form factors which are given by
\begin{equation}
\begin{array}{l}
  \sqrt{6} f_{A} = 
       -(w + 1) \tau \\
    \hspace{1.2cm}
     - \varepsilon_b (w - 1)[(\bar{\Lambda}' + \bar{\Lambda}) \tau
         - (2 w + 1) \tau_1 - \tau_2] \\
    \hspace{1.2cm}
     - \varepsilon_c [ 4 (w \bar{\Lambda}' - \bar{\Lambda}) \tau
         - 3(w - 1)(\tau_1 - \tau_2)] \\
  \sqrt{6} f_{V_1} = 
       (1 - w^2) \tau \\
    \hspace{1.2cm}
     - \varepsilon_b (w^2 - 1)[(\bar{\Lambda}' + \bar{\Lambda}) \tau
         - (2 w + 1) \tau_1 - \tau_2 ] \\
    \hspace{1.2cm}
     - \varepsilon_c [ 4 (w + 1)(w \bar{\Lambda}' - \bar{\Lambda}) \tau
         - 3(w^2 - 1)(\tau_1 - \tau_2) ] \\
  \sqrt{6} f_{V_2} = 
     -3 \tau
     - 3\varepsilon_b [(\bar{\Lambda}' + \bar{\Lambda}) \tau
         - (2 w + 1) \tau_1 - \tau_2 ] \\
    \hspace{1.2cm}
     - \varepsilon_c [(4w - 1)\tau_1 + 5\tau_2 ] \\
  \sqrt{6} f_{V_3} = 
    (w - 2) \tau \\
    \hspace{1.2cm}
     + \varepsilon_b (2 + w)[(\bar{\Lambda}' + \bar{\Lambda}) \tau
         - (2 w + 1) \tau_1 - \tau_2] \\
    \hspace{1.2cm}
     + \varepsilon_c [ 4 (w \bar{\Lambda}' - \bar{\Lambda}) \tau
         + (2 + w)\tau_1 + (2 + 3w)\tau_2 ]. 
\end{array} 
\end{equation}
Here $\tau$ is the leading Isgur-Wise function, which is assumed to be a linear form~\cite{ref:LLSW}
\begin{equation}
  \tau(w) = \tau(1)[1+\hat{\tau}'(w-1)].
\end{equation}
Uncertainty in first order expansion of Isgur-Wise function
is parameterized in $\tau_1$ and $\tau_2$. 
In approximation B$_1$ one sets
\begin{equation}
  \tau_1 = 0,
  \hspace{0.3cm}
  \tau_2 = 0 ,
\end{equation}
while in approximation B$_2$ one takes
\begin{equation}
  \tau_1 = \bar{\Lambda}\tau,
  \hspace{0.3cm}
  \tau_2 = -\bar{\Lambda}'\tau.
\end{equation}

\subsubsection{$\Bb\to D_2^{*} \ell \nub$}
\noindent The differential decay rate is given by
\begin{equation}
  \frac{d^2 \Gamma_{D_2^{*}}}{d w\, d\cos\theta}
   =  \Gamma_z 
    r^3 (w^2 - 1)^{3/2} \hspace{0.1cm}
    \frac{1}{2} \hspace{0.1cm}
    \mathcal{I}_{D_2^{*}} (w, \theta),
\end{equation}
where
\begin{eqnarray}
   \mathcal{I}_{D_2^{*}}(w, \theta) \hspace{6.2cm}\nonumber\\ 
   = \frac{4}{3} (1-\cos^2\theta) 
      [(w - r) k_{A_1} + (w^2 - 1)(k_{A_3} + r k_{A_2})]^2 
          \nonumber  \\
   + (1 - 2 r w + r^2)
      \lbrack(1 + \cos^2\theta)(k_{A_1}^2 + (w^2 - 1)k_{V}^2) \nonumber  \\
       -4\cos\theta \sqrt{w^2 - 1} k_{A_1} k_{V}\rbrack \hspace{0.5cm}
\end{eqnarray}
and $k_{V}(w)$, $k_{A_1}(w)$, $k_{A_2}(w)$ and $k_{A_3}(w)$ are form factors which are given by
\begin{equation}
\begin{array}{l}
  k_V = -\tau 
       -\varepsilon_b
         [(\bar{\Lambda}' + \bar{\Lambda})\tau 
	             - (2w+1)\tau_1 - \tau_2] \\
   \hspace{0.8cm}
       - \varepsilon_c 
         (\tau_1 - \tau_2) \\
  k_{A_1} =  -(1+w)\tau \\
   \hspace{0.8cm}
       -\varepsilon_b 
         (w-1)[(\bar{\Lambda}' + \bar{\Lambda})\tau 
	             - (2w+1)\tau_1 - \tau_2]  \\
   \hspace{0.8cm}
     - \varepsilon_c 
         (w-1)(\tau_1 - \tau_2) \\
  k_{A_2} = -2\varepsilon_c \tau_1 \\
  k_{A_3} = \tau 
       +\varepsilon_b
         [(\bar{\Lambda}' + \bar{\Lambda})\tau 
	             - (2w+1)\tau_1 - \tau_2] \\
   \hspace{0.8cm}
       - \varepsilon_c (\tau_1 + \tau_2). 
\end{array}
\end{equation}

\subsubsection{$\Bb\to D_0^{*} \ell \nub$}
\noindent The differential decay rate is given by
\begin{equation}
  \frac{d^2 \Gamma_{D_0^*}}{d w\, d\cos\theta}
   = \Gamma_z 
    r^3 (w^2 - 1)^{3/2} \hspace{0.1cm}
    \mathcal{I}_{D_0^*}(w, \theta),
\end{equation}
where
\begin{equation}
    \mathcal{I}_{D_0^*}(w, \theta)
    = (1-\cos^2\theta)[(1 + r) g_+ - (1 - r) g_-]^2
\end{equation}
and $g_+(w)$ and $g_-(w)$ are form factors which are given by
\begin{equation}
\begin{array}{l}
  g_+ = 
    \varepsilon_c \left[
      2(w-1)\zeta_1 
      - 3\zeta \frac{w \bar{\Lambda}^* - \bar{\Lambda}}{w+1}
    \right] \\
   \hspace{0.8cm}
    -\varepsilon_b \left[
      \frac{\bar{\Lambda}^* (2w+1) - \bar{\Lambda} (w+2)}{w+1} \zeta
      -2(w-1)\zeta_1
    \right] \\
  g_- =  \zeta
\end{array}
\end{equation}
with
\begin{equation}
  \zeta(w) = \frac{w+1}{\sqrt{3}}\tau(w).
\end{equation}
In approximation B$_1$ one uses
\begin{equation}
  \zeta_1 = 0,
  \hspace{0.3cm}
  \zeta_2 = 0 ,
\end{equation}
while in approximation B$_2$ one takes
\begin{equation}
  \zeta_1 = \bar{\Lambda}\zeta, 
  \hspace{0.3cm}
  \zeta_2 = -\bar{\Lambda}^*\zeta.
\end{equation}

\subsubsection{$\Bb\to D_1' \ell \nub$}
\noindent The differential decay rate is given by
\begin{equation}
  \frac{d^2 \Gamma_{D_1'}}{d w\, d\cos\theta}
   =  \Gamma_z 
    r^3 (w^2 - 1)^{1/2} \hspace{0.1cm}
   \mathcal{I}_{D_1'} (w, \theta),
\end{equation}
where
\begin{eqnarray}
   \mathcal{I}_{D_1'}(w, \theta)  \hspace{6.2cm}\nonumber\\
   = (1-\cos^2\theta) [(w - r) g_{V_1} 
                     + (w^2 - 1)(g_{V_3} + r g_{V_2})]^2 \nonumber\\
   + (1 - 2 r w + r^{2})
      [(1 + \cos^2\theta)(g_{V_1}^2 + (w^2 - 1)g_{A}^2) \nonumber\\
       -4\cos\theta \sqrt{w^2 - 1} g_{V_1} g_{A}] \hspace{0.3cm}
\end{eqnarray}
and $g_{V_1}(w)$, $g_{V_2}(w)$, $g_{V_3}(w)$ and $g_{A}(w)$ are form factors which are given by
\begin{equation}
\begin{array}{l}
  g_{A} = \zeta
     + \varepsilon_c
      \left[\frac{w\bar{\Lambda}^* - \bar{\Lambda}}{w+1} \zeta
      \right] \\
   \hspace{0.8cm}
      - \varepsilon_b
      \left[
	\frac{\bar{\Lambda}^* (2w+1) - \bar{\Lambda} (w+2)}{w+1} \zeta
	-2(w-1)\zeta_1
      \right] \\
  g_{V_1} = (w-1)\zeta + \varepsilon_c
	(w\bar{\Lambda}^* - \bar{\Lambda})\zeta \\
   \hspace{0.8cm}
     - \varepsilon_b
      \left[
	(\bar{\Lambda}^* (2w+1) - \bar{\Lambda} (w+2))\zeta
	-2(w^2-1)\zeta_1 
      \right] \\
  g_{V_2} = 2\varepsilon_c \zeta_1 \\
  g_{V_3} = -\zeta - \varepsilon_c 
      \left[\frac{w\bar{\Lambda}^* - \bar{\Lambda}}{w+1} \zeta
           + 2\zeta_1   \right] \\
   \hspace{0.8cm}
      + \varepsilon_b
      \left[
	\frac{\bar{\Lambda}^* (2w+1) - \bar{\Lambda} (w+2)}{w+1} \zeta
	-2(w-1)\zeta_1 
      \right]. 
\end{array}
\end{equation}

\vfill

\end{document}